\pdfoutput=1
\documentclass[12pt,a4paper]{article}

\usepackage{ifthen} 
\newboolean{pdflatex}
\setboolean{pdflatex}{true} 

\newboolean{articletitles}
\setboolean{articletitles}{true} 

\newboolean{uprightparticles}
\setboolean{uprightparticles}{false} 

\newboolean{inbibliography}
\setboolean{inbibliography}{false} 

\def\paperauthors{LHCb collaboration} 
\def\paperasciititle{} 
\def\papertitle{Search for the baryon- and lepton-number violating decays\\ $B^0\to p\mu^-$ and $B^0_s\to p\mu^-$} 
\def\paperkeywords{{High Energy Physics}, {LHCb}} 
\def\papercopyright{CERN on behalf of the LHCb collaboration}
\def\paperlicence{CC BY 4.0 licence}
\def\paperlicenceurl{https://creativecommons.org/licenses/by/4.0/}
\usepackage{booktabs}

\usepackage[top=1in, bottom=1.25in, left=1in, right=1in]{geometry}

\usepackage{microtype}
\usepackage{lineno}  
\usepackage{xspace} 
\usepackage{caption} 

\usepackage{graphicx}  
\usepackage{color}
\usepackage{colortbl}
\graphicspath{{./figs/}} 

\usepackage{amsmath} 
\usepackage{amssymb}
\usepackage{amsfonts}
\usepackage{upgreek} 

\newcommand*\patchAmsMathEnvironmentForLineno[1]{%
\expandafter\let\csname old#1\expandafter\endcsname\csname #1\endcsname
\expandafter\let\csname oldend#1\expandafter\endcsname\csname
end#1\endcsname
 \renewenvironment{#1}%
   {\linenomath\csname old#1\endcsname}%
   {\csname oldend#1\endcsname\endlinenomath}%
}
\newcommand*\patchBothAmsMathEnvironmentsForLineno[1]{%
  \patchAmsMathEnvironmentForLineno{#1}%
  \patchAmsMathEnvironmentForLineno{#1*}%
}
\AtBeginDocument{%
\patchBothAmsMathEnvironmentsForLineno{equation}%
\patchBothAmsMathEnvironmentsForLineno{align}%
\patchBothAmsMathEnvironmentsForLineno{flalign}%
\patchBothAmsMathEnvironmentsForLineno{alignat}%
\patchBothAmsMathEnvironmentsForLineno{gather}%
\patchBothAmsMathEnvironmentsForLineno{multline}%
\patchBothAmsMathEnvironmentsForLineno{eqnarray}%
}

\usepackage{hyperxmp}

\usepackage[pdftex,
            pdfauthor={\paperauthors},
            pdftitle={\paperasciititle},
            pdfkeywords={\paperkeywords},
            pdfcopyright={Copyright (C) \papercopyright},
            pdflicenseurl={\paperlicenceurl}]{hyperref}

\usepackage[colorinlistoftodos,textsize=scriptsize]{todonotes}

\usepackage[bottom,flushmargin,hang,multiple]{footmisc}

\usepackage[all]{hypcap} 

\usepackage{xspace} 
\usepackage{upgreek}

\def\lhcb   {\mbox{LHCb}\xspace}

\def\MagUp {\mbox{\em Mag\kern -0.05em Up}\xspace}

\ifthenelse{\boolean{uprightparticles}}%
{

 \def\Pmu         {\ensuremath{\upmu}\xspace}                 
 \def\Pnu         {\ensuremath{\upnu}\xspace}                 
                  
 \def\Ppi         {\ensuremath{\uppi}\xspace}

 \def\Ppsi        {\ensuremath{\uppsi}\xspace}

 \def\PDelta      {\ensuremath{\Delta}\xspace}                 
 \def\PXi         {\ensuremath{\Xi}\xspace}                 
 \def\PLambda     {\ensuremath{\Lambda}\xspace}                 
 \def\PSigma      {\ensuremath{\Sigma}\xspace}                 
 \def\POmega      {\ensuremath{\Omega}\xspace}                 
 \def\PUpsilon    {\ensuremath{\Upsilon}\xspace}
 \let\oldPi\Pi
 \def\PPi         {\ensuremath{\oldPi}\xspace}

 \def\PB      {\ensuremath{\mathrm{B}}\xspace}                 
                  
 \def\PD      {\ensuremath{\mathrm{D}}\xspace}

 \def\PJ      {\ensuremath{\mathrm{J}}\xspace}                 
 \def\PK      {\ensuremath{\mathrm{K}}\xspace}

 \def\Pb      {\ensuremath{\mathrm{b}}\xspace}                 
 \def\Pc      {\ensuremath{\mathrm{c}}\xspace}

 \def\Pi      {\ensuremath{\mathrm{i}}\xspace}

 \def\Pp      {\ensuremath{\mathrm{p}}\xspace}

 \def\Ps      {\ensuremath{\mathrm{s}}\xspace}                 
                  
 \def\Pu      {\ensuremath{\mathrm{u}}\xspace}

 \def\thebaroffset{0.0em}
}
{

 \def\Pmu         {\ensuremath{\mu}\xspace}                 
 \def\Pnu         {\ensuremath{\nu}\xspace}                 
                  
 \def\Ppi         {\ensuremath{\pi}\xspace}

 \def\Ppsi        {\ensuremath{\psi}\xspace}                 
                  
 \mathchardef\PDelta="7101
 \mathchardef\PXi="7104
 \mathchardef\PLambda="7103
 \mathchardef\PSigma="7106
 \mathchardef\POmega="710A
 \mathchardef\PUpsilon="7107
 \mathchardef\PPi="7105
                  
 \def\PB      {\ensuremath{B}\xspace}                 
                  
 \def\PD      {\ensuremath{D}\xspace}

 \def\PJ      {\ensuremath{J}\xspace}                 
 \def\PK      {\ensuremath{K}\xspace}

 \def\Pb      {\ensuremath{b}\xspace}                 
 \def\Pc      {\ensuremath{c}\xspace}

 \def\Pi      {\ensuremath{i}\xspace}

 \def\Pp      {\ensuremath{p}\xspace}

 \def\Ps      {\ensuremath{s}\xspace}                 
                  
 \def\Pu      {\ensuremath{u}\xspace}

 \def\thebaroffset{0.18em}
}
\newcommand{\offsetoverline}[2][\thebaroffset]{\kern #1\overline{\kern -#1 #2}}%

\makeatletter
\ifcase \@ptsize \relax
  \newcommand{\miniscule}{\@setfontsize\miniscule{4}{5}}
\or
  \newcommand{\miniscule}{\@setfontsize\miniscule{5}{6}}
\or
  \newcommand{\miniscule}{\@setfontsize\miniscule{5}{6}}
\fi
\makeatother

\DeclareRobustCommand{\optbar}[1]{\shortstack{{\miniscule (\rule[.5ex]{1.25em}{.18mm})}
  \\ [-.7ex] $#1$}}

\def\muon       {{\ensuremath{\Pmu}}\xspace}
\def\mup        {{\ensuremath{\Pmu^+}}\xspace}
\def\mun        {{\ensuremath{\Pmu^-}}\xspace} 

\def\mumu       {{\ensuremath{\Pmu^+\Pmu^-}}\xspace}

\def\neu        {{\ensuremath{\Pnu}}\xspace}

\def\neum       {{\ensuremath{\neu_\mu}}\xspace}

\def\uquark    {{\ensuremath{\Pu}}\xspace}

\def\squark    {{\ensuremath{\Ps}}\xspace}

\def\cquark    {{\ensuremath{\Pc}}\xspace}

\def\bquark    {{\ensuremath{\Pb}}\xspace}

\def\pion   {{\ensuremath{\Ppi}}\xspace}

\def\pip    {{\ensuremath{\pion^+}}\xspace}
\def\pim    {{\ensuremath{\pion^-}}\xspace}

\def\kaon    {{\ensuremath{\PK}}\xspace}

\def\KorKbar {\kern \thebaroffset\optbar{\kern -\thebaroffset \PK}{}\xspace}

\def\Kp      {{\ensuremath{\kaon^+}}\xspace}
\def\Km      {{\ensuremath{\kaon^-}}\xspace}

\def\D       {{\ensuremath{\PD}}\xspace}

\def\DorDbar {\kern \thebaroffset\optbar{\kern -\thebaroffset \PD}\xspace}
\def\Dz      {{\ensuremath{\D^0}}\xspace}

\def\Dp      {{\ensuremath{\D^+}}\xspace}
\def\Dm      {{\ensuremath{\D^-}}\xspace}

\def\DpDm    {\ensuremath{\Dp {\kern -0.16em \Dm}}\xspace}

\def\B       {{\ensuremath{\PB}}\xspace}

\def\BorBbar {\kern \thebaroffset\optbar{\kern -\thebaroffset \PB}\xspace}

\def\Bd      {{\ensuremath{\B^0}}\xspace}

\def\BdorBdbar {\kern \thebaroffset\optbar{\kern -\thebaroffset \Bd}\xspace}
\def\Bu      {{\ensuremath{\B^+}}\xspace}

\def\Bs      {{\ensuremath{\B^0_\squark}}\xspace}

\def\BsorBsbar {\kern \thebaroffset\optbar{\kern -\thebaroffset \Bs}\xspace}

\def\Bds     {{\ensuremath{\B_{(\squark)}^0}}\xspace}

\def\jpsi     {{\ensuremath{{\PJ\mskip -3mu/\mskip -2mu\Ppsi}}}\xspace}

\def\Y#1S{\ensuremath{\PUpsilon{(#1S)}}\xspace}

\def\proton      {{\ensuremath{\Pp}}\xspace}
\def\antiproton  {{\ensuremath{\overline \proton}}\xspace}

\def\Lz          {{\ensuremath{\PLambda}}\xspace}

\def\LorLbar     {\kern \thebaroffset\optbar{\kern -\thebaroffset \PLambda}\xspace}

\def\Lb           {{\ensuremath{\Lz^0_\bquark}}\xspace}

\def\BF         {{\ensuremath{\mathcal{B}}}\xspace}
\def\BR         {\BF}

\newcommand{\decay}[2]{\ensuremath{#1\!\to #2}\xspace} 

\def\to                 {\ensuremath{\rightarrow}\xspace}

\def\BdToKpi      {\decay{\Bd}{\Kp\pim}}
\def\BsToKK       {\decay{\Bs}{\Kp\Km}}

\def\AT#1     {\ensuremath{A_{\mathrm{T}}^{#1}}\xspace}           

\def\C#1      {\ensuremath{\mathcal{C}_{#1}}\xspace}                       
\def\Cp#1     {\ensuremath{\mathcal{C}_{#1}^{'}}\xspace}                    
\def\Ceff#1   {\ensuremath{\mathcal{C}_{#1}^{\mathrm{(eff)}}}\xspace}        
\def\Cpeff#1  {\ensuremath{\mathcal{C}_{#1}^{'\mathrm{(eff)}}}\xspace}       
\def\Ope#1    {\ensuremath{\mathcal{O}_{#1}}\xspace}                       
\def\Opep#1   {\ensuremath{\mathcal{O}_{#1}^{'}}\xspace}                    

\newcommand{\aunit}[1]{\ensuremath{\text{\,#1}}}       

\newcommand{\tev}{\aunit{Te\kern -0.1em V}\xspace}
\newcommand{\gev}{\aunit{Ge\kern -0.1em V}\xspace}
\newcommand{\mev}{\aunit{Me\kern -0.1em V}\xspace}
\newcommand{\kev}{\aunit{ke\kern -0.1em V}\xspace}
\newcommand{\ev}{\aunit{e\kern -0.1em V}\xspace}
 
\newcommand{\mevc}{\ensuremath{\aunit{Me\kern -0.1em V\!/}c}\xspace}
\newcommand{\gevc}{\ensuremath{\aunit{Ge\kern -0.1em V\!/}c}\xspace}
\newcommand{\mevcc}{\ensuremath{\aunit{Me\kern -0.1em V\!/}c^2}\xspace}
\newcommand{\gevcc}{\ensuremath{\aunit{Ge\kern -0.1em V\!/}c^2}\xspace}

\def\fb   {\ensuremath{\aunit{fb}}\xspace}
\def\invfb   {\ensuremath{\fb^{-1}}\xspace}

\newcommand{\chisq}{\ensuremath{\chi^2}\xspace}

\newcommand{\chisqip}{\ensuremath{\chi^2_{\text{IP}}}\xspace}

\newcommand{\chisqvtx}{\ensuremath{\chi^2_{\text{vtx}}}\xspace}

\def\gsim{{~\raise.15em\hbox{$>$}\kern-.85em
          \lower.35em\hbox{$\sim$}~}\xspace}
\def\lsim{{~\raise.15em\hbox{$<$}\kern-.85em
          \lower.35em\hbox{$\sim$}~}\xspace}

\def\sqs   {\ensuremath{\protect\sqrt{s}}\xspace}

\def\pt         {\ensuremath{p_{\mathrm{T}}}\xspace}

\def\ptot       {\ensuremath{p}\xspace}

\def\evtgen     {\mbox{\textsc{EvtGen}}\xspace}

\def\geant      {\mbox{\textsc{Geant4}}\xspace}

\def\photos     {\mbox{\textsc{Photos}}\xspace}

\def\pythia     {\mbox{\textsc{Pythia}}\xspace}

\def\tell1  {TELL1\xspace}
\def\ukl1   {UKL1\xspace}

\newcommand{\lhcborcid}[1]{\href{https://orcid.org/#1}{\hspace*{0.1em}\raisebox{-0.45ex}{\includegraphics[width=1em]{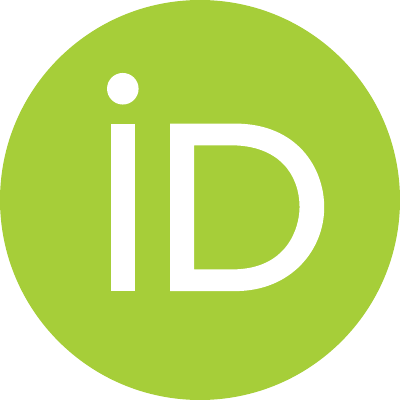}}}}

\usepackage{cite} 
\usepackage{mciteplus}

\def\Bds     {{\ensuremath{\B^0_{(\squark)}}}\xspace}

\def\LbTopK       {\decay{\Lb}{\proton\Km}}
\def\LbToppi      {\decay{\Lb}{\proton\pim}}
\def\LbTopmunu    {\decay{\Lb}{\proton\mun\overline{\nu}_{\mu}}}
\def\LbTopXmunu    {\decay{\Lb}{\proton X\mun\overline{\nu}_{\mu}}}

\def\BdTopimunu   {\decay{\Bd}{\pim\mup\neum}}
\def\BsToKmunu    {\decay{\Bs}{\Km\mup\neum}}

\def\jpsimumu    {{\ensuremath{{\PJ\mskip -3mu/\mskip -2mu\Ppsi\mskip 2mu(\mumu)}}}\xspace}
\def\JpsiTomumu  {\decay{\jpsi}{\mup\mun}}
\def\BuToJPsiK   {\decay{\Bu}{\jpsi\Kp}}

\def\BuToJPsimuK  {\decay{\Bu}{\jpsimumu\Kp}}

\def\BsToKpi      {\decay{\Bs}{\Kp\pim}}

\def\DToKpi      {\decay{\Dz}{\Km\pip}}

\def\BdTopmu  {\decay{\Bd}{\proton\mun}}

\def\BsTopmu  {\decay{\Bs}{\proton\mun}}
\def\BdsTopmu {\decay{\Bds}{\proton\mun}}

\usepackage{longtable} 
\begin{document}

\renewcommand{\thefootnote}{\fnsymbol{footnote}}
\setcounter{footnote}{1}


\begin{titlepage}
\pagenumbering{roman}

\vspace*{-1.5cm}
\centerline{\large EUROPEAN ORGANIZATION FOR NUCLEAR RESEARCH (CERN)}
\vspace*{1.5cm}
\noindent
\begin{tabular*}{\linewidth}{lc@{\extracolsep{\fill}}r@{\extracolsep{0pt}}}
\ifthenelse{\boolean{pdflatex}}

\\
 & & CERN-EP-2022-195 \\  
 & & LHCb-PAPER-2022-022 \\  
 & & \today \\ 
 & & \\
\end{tabular*}

\vspace*{2.5cm}

{\normalfont\bfseries\boldmath\huge
\begin{center}
  \papertitle 
\end{center}
}

\vspace*{2.0cm}

\begin{center}
\paperauthors\footnote{Authors are listed at the end of this paper.}
\end{center}

\vspace{\fill}

\begin{abstract}
  \noindent
  A search for the baryon- and lepton-number violating decays \BdTopmu and \mbox{\BsTopmu} is performed at the \lhcb experiment using data collected in proton-proton collisions at \sqs = 7, 8 and 13 \tev, corresponding to integrated luminosities of 1, 2 and 6 \invfb, respectively. No significant signal for \BdTopmu and \BsTopmu decays is found and the upper limits on the branching fractions are determined to be \mbox{$\BR(\BdTopmu)<2.6\,(3.1)\times10^{-9}$} and \mbox{$\BR(\BsTopmu)<12.1\,(14.0)\times10^{-9}$}, respectively, at 90\% (95\%) confidence level. 
  These are the first limits on these decays to date.
\end{abstract}

\vspace*{2.0cm}

\begin{center}
Published in Phys. Rev. D108 (2023) 012021
\end{center}

\vspace{\fill}

{\footnotesize 
\centerline{\copyright~\papercopyright. \href{\paperlicenceurl}{\paperlicence}.}}
\vspace*{2mm}

\end{titlepage}


\newpage
\setcounter{page}{2}
\mbox{~}


\renewcommand{\thefootnote}{\arabic{footnote}}
\setcounter{footnote}{0}

\cleardoublepage


\pagestyle{plain} 
\setcounter{page}{1}
\pagenumbering{arabic}

\section{Introduction}
\label{sec:introduction}

Despite the tremendous success of the Standard Model (SM) of particle physics in the past few decades, there are still some important questions that are not answered. One of them is the
observation of the matter-antimatter asymmetry in the universe, which brings a serious challenge to our understanding of nature. In 1967, Andrei Sakharov proposed three necessary conditions that must be satisfied to produce such a large matter-antimatter asymmetry, one of which is baryon number violation (BNV)~\cite{Sakharov:1967dj}. Various extensions of the SM that include BNV processes, known as Grand Unified Theory (GUT) models, have been proposed, such as SU(5)~\cite{PhysRevLett.32.438}, SO(10)~\cite{FRITZSCH1975193}, E$_6$~\cite{Gursey:1975ki} and flipped SU(5)~\cite{Barr:1981qv,Derendinger:1983aj}. 
These GUT models usually require two hypothetical gauge bosons, $X$ and $Y$, with electric charges of $\pm \frac{4}{3}e$ and $\pm \frac{1}{3}e$, that couple quarks to leptons and lead to both BNV and lepton number violation (LNV). 

\begin{figure}
\centering
\includegraphics[scale=0.7]{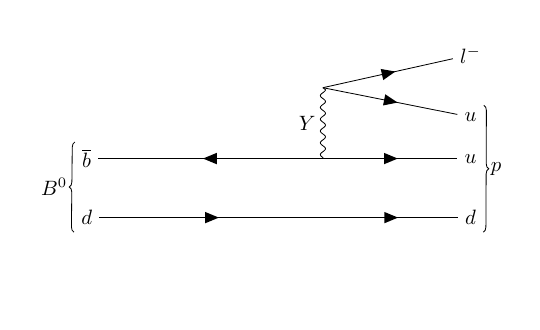}\includegraphics[scale=0.7]{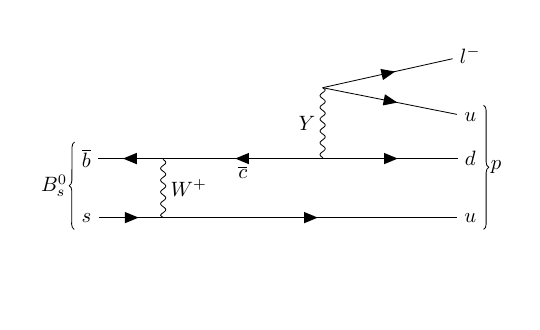}
\includegraphics[scale=0.7]{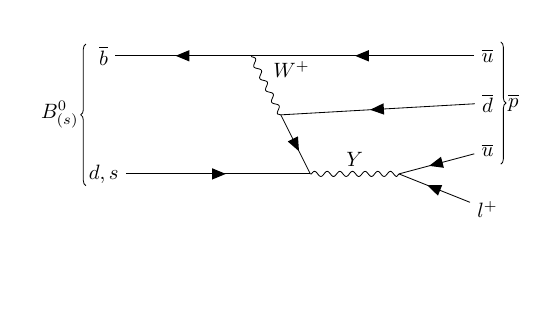}

\caption{Hypothetical Feynman diagrams of $\Bds \rightarrow \proton \ell^{-}$ and $\antiproton\ell^{+}$ mediated by a hypothetical $Y$ boson.}

\label{fig:B2uulprocess}
\end{figure}

Proton decay is a BNV process of the lightest baryon: none of its decay modes, although predicted by many GUT models, have been observed~\cite{ParticleDataGroup:2020ssz}. However, since proton decay only involves first-generation quarks, experimental searches for BNV decays of heavy-flavour hadrons are very important and represent additional probes for new physics effects. Various BNV processes have been searched for in $\tau$, $\Lz$, $\D$, $J/\psi$, and $\B$
decays by the CLEO~\cite{CLEO:1999emi}, CLAS~\cite{McCracken:2015coa}, BESIII~\cite{BESIII:2022svy,BESIII:2021krj,BESIII:2019udi,BESIII:2018cls} and BABAR~\cite{BaBar:2011yks}
experiments, but no evidence has been found so far.  The large data samples accumulated by the \lhcb experiment are expected to lead to the best sensitivity for investigating BNV decays of $\B$ mesons.
The $\Bds \rightarrow \proton \ell^{-}$ decay modes, with possible hypothetical Feynman diagrams shown in Fig.~\ref{fig:B2uulprocess}, are relevant BNV and LNV processes that are forbidden in the SM with a prediction of $\mbox \BR(\bar{\bquark} \rightarrow \uquark\uquark \ell^{-})<2.4\times10^{-27}$~\cite{Hou:2005iu} derived from the constraint of proton stability.

 In this paper, we present the first search for \BdTopmu and \BsTopmu decays\footnote{\BdsTopmu represents \BdsTopmu and $\antiproton\mup$ and the inclusion of charge-conjugate processes is implied throughout this paper, unless otherwise noted.} using proton-proton (\proton\proton) collisions collected with the \lhcb detector and corresponding to an integrated  luminosity of 1~\invfb at a center-of-mass energy \sqs = 7~\tev, 2~\invfb at \sqs = 8~\tev and 6~\invfb at \sqs = 13~\tev. The first two data sets are referred to as Run 1 and the latter as Run 2. Two normalisation channels are used: the \BdToKpi decay, which has a similar topology to that of the signal, and the \BuToJPsimuK decay, due to its high abundance and the high purity that can be achieved at the LHCb experiment. The \BdsTopmu candidates with $p\muon^{-}$ pair invariant mass $m_{p\muon^{-}}$ in the range [5067, 5667]\mevcc are selected. To avoid potential biases, \BdsTopmu candidates in the signal region, $m_{p\muon^{-}}$ $\in$ [5217, 5457]\mevcc, were not examined until the selection and fitting procedure were finalised. 

\section{Detector and simulation}
The \lhcb detector~\cite{LHCb:2008vvz,LHCb:2014set} is a single-arm forward spectrometer covering the pseudorapidity range 2 $< \eta <$ 5, designed for the study of particles containing \bquark or \cquark quarks.
The detector includes a high-precision tracking system consisting of a silicon-strip vertex
detector surrounding the \proton\proton interaction region, a large-area silicon-strip detector located
upstream of a dipole magnet with a bending power of about 4 Tm, and three stations of
silicon-strip detectors and straw drift tubes placed downstream of the magnet. The tracking system provides a measurement of momentum, \ptot, of charged particles with a relative
uncertainty that varies from 0.5\% at low momentum to 1.0\% at 200\gevc. The minimum
distance of a track to a primary vertex (PV), the impact parameter (IP), is measured with
a resolution of (15 + 29/\pt) \textup{$\upmu$}m, where \pt is the component of the momentum transverse
to the beam, in \gevc. Different types of charged hadrons are distinguished using information from two ring-imaging Cherenkov detectors. Photons, electrons and hadrons are
identified by a calorimeter system consisting of scintillating-pad and preshower detectors,
an electromagnetic calorimeter and a hadronic calorimeter. Muons are identified by a
system composed of alternating layers of iron and multiwire proportional chambers.

The online event selection is performed by a trigger~\cite{Aaij:2012me}, which consists of a hardware stage,
based on information from the muon and calorimeter systems, followed by a software stage
that applies a full reconstruction of the event. At the hardware stage, the presence of a muon with high \pt is required, while at the software level both tracks are used. A first software stage requires the presence of at
least one track with high \pt that is well separated from the PV. It is followed by a second stage that 
requires a two-track secondary vertex with significant displacement from any
PV, and a multivariate algorithm~\cite{Gligorov:2012qt, Likhomanenko:2015aba} is used for the identification of secondary vertices consistent with the decay of a $b$ hadron.

Simulated samples of signal and background channels are used to evaluate geometrical, reconstruction and selection efficiencies, to train multivariate classifiers and to determine the shapes of invariant mass distributions of both signal and background contributions. In the simulation, \proton\proton collisions are generated using \pythia~\cite{Sjostrand:2007gs} with a specific \lhcb configuration~\cite{LHCb:2011dpk}. Decays of hadronic particles are described by \evtgen~\cite{Lange:2001uf}, in which final-state radiation is generated using \photos~\cite{Golonka:2005pn}. The signal events are generated by assuming the $B^0_{(s)}$ decays do not produce any preferred polarization of the proton. The interaction of the generated particles with the detector, and its response, are simulated using the \geant toolkit~\cite{Allison:2006ve,GEANT4:2002zbu} as described in Ref.~\cite{Clemencic:2011zza}. 

\section{Selection}
The \BdsTopmu candidates are reconstructed by combining two oppositely charged tracks with transverse momentum in the range 0.25 $< \pt <$ 40 \gevc, momentum \ptot $<$ 500 \gevc, as in Ref.\cite{LHCb:2021awg}. The tracks are also required to have a good track fit quality, with a track fit \chisq per degree of freedom smaller than three.  Only candidate tracks with \chisqip $>$ 25 for any PV are selected, where \chisqip is defined as the difference between the vertex-fit \chisq of the PV formed with and without the particle in question. The distance of closest approach between the two tracks is required to be below 0.1 mm. The proton candidates are required to be in the geometric acceptance of the LHCb muon stations to mimic the selection of the muon pair of the normalisation channel \BuToJPsimuK. This requirement has a negligible effect on the signal efficiency. The two candidate tracks are required to form a secondary vertex with a vertex-fit \chisq per degree of freedom (\chisqvtx) smaller than nine. Furthermore, the resulting \Bds candidates must have a decay time less than nine times the \Bs lifetime, \chisqip $<$ 25 with respect to the PV for which the \chisqip is minimal 
and \pt $>$ 0.5\gevc, to suppress background from random combinations of tracks originating from the PV.

Particles forming the \BdsTopmu candidates are required to be well identified as a proton and a muon~\cite{Archilli:2013npa}, using information from the Cherenkov detectors, the calorimeters
and the muon stations. The stringent requirements on the \proton and \muon candidates retain more than 60\%  of the signal candidates and eliminate more than 99\% of the background candidates from decays with misidentified particles, particularly the two-body peaking backgrounds, but have almost no discriminating power on the semileptonic decay  \mbox{\LbTopmunu}, which has the same visible final state particles as the signal channels. This decay represents the main physics background in this analysis. The \LbTopXmunu decays with any additional unreconstructed particles ($X$) are expected to have negligible contribution in the fit region of $p\mu^-$ mass. 

For the normalization channels, the selection for \BdToKpi candidates is the same as for the \BdsTopmu channel, except for the particle identification (PID) criteria. 
Similarly, the \BuToJPsimuK candidate selection is also kept as similar as possible, applying the same selection used for the signal to the dimuon pair from the \jpsi decay. Additionally, loose quality requirements are applied on the \jpsi vertex, \chisqvtx $<$ 9. Finally, a 100 \mevcc mass window around the known \Bu mass and a 60 \mevcc mass window around the known \jpsi mass\cite{PDG2020} are used. 

\section{MLP training and calibration}
\label{subsec:MLPCalibration}
\begin{figure}
\centering
\includegraphics[scale=0.35]{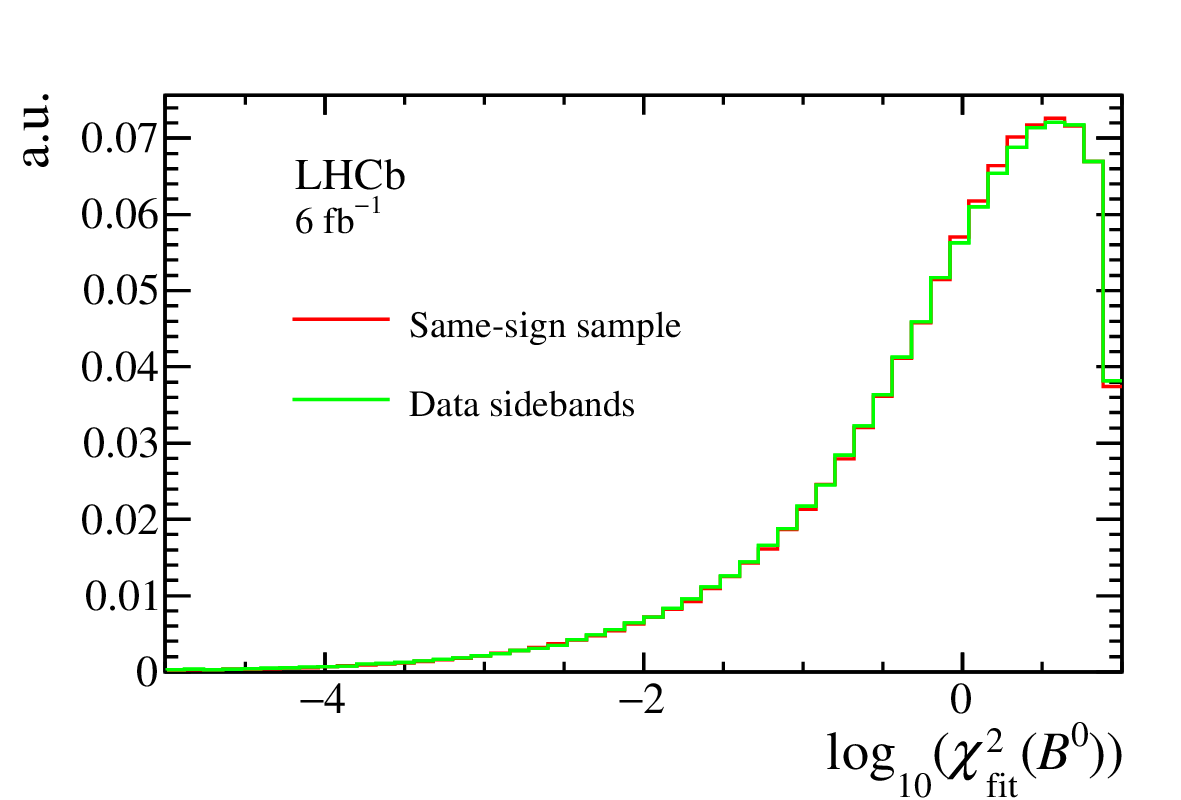}\includegraphics[scale=0.35]{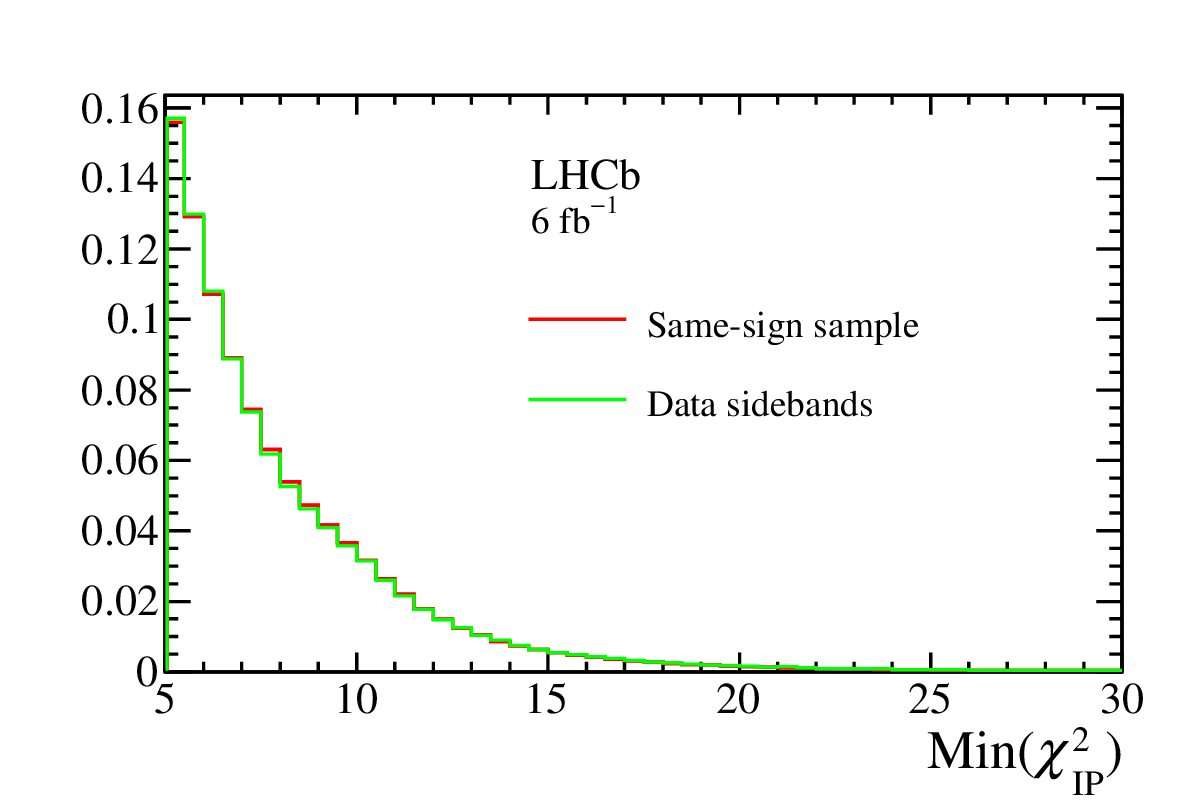}
\caption{Comparison of the distributions of the \Bd vertex fit \chisq ($\chi^2_{\mathrm{fit}}(B^0)$) and the minimum \chisqip of the two tracks in the final state ($\mathrm{Min}(\chi^2_{\mathrm{IP}})$), for the same-sign sample and data sidebands in Run 2.}
\label{fig:Comp}
\end{figure}

A Multilayer perceptron (MLP)~\cite{TMVA4} classifier implemented in the TMVA toolkit~\cite{Hocker:2007ht} is used to separate the \BdsTopmu signal from the combinatorial background, which arises from random combinations of tracks. The classifier is trained using a sample of simulated \BdTopmu events to describe the signal, and a data sample of same-sign $p\muon^{+}$ candidates to describe the combinatorial background. The same-sign sample reproduces the distributions of the MLP input variables from the data sidebands ($5067 < m_{p\muon^{-}} < 5100$ \mevcc or $5500 < m_{p\muon^{-}} < 5667$\mevcc). The following input variables are used: the minimum \chisqip of the two tracks in the final state and the \chisqip of the \Bd candidate, the vertex fit \chisq of the \Bd decay and its displacement from the production vertex, the \Bd transverse momentum and its proper decay time, the distance of closest approach between the two tracks in the final state, the difference of pseudorapidity between the two final-state tracks, the angular difference between the direction of \Bd momentum and the direction defined by the secondary and primary vertices, and the angular difference between the direction of the \muon momentum and the vector perpendicular to the \Bd momentum and the beam axis in the \Bd rest frame. Distributions of two classifier input variables, the vertex fit \chisq of the \Bd decay and the minimum \chisqip of the two tracks in the final state, for the same-sign sample and the data sidebands in Run 2 are compared in Fig.~\ref{fig:Comp}.

The response of the MLP classifier is constructed to be uniform in the range [0,1] for signal 
after the full selection without PID requirements. For background, the MLP response peaks near zero. Its linear correlation with the $p\muon^{-}$ pair mass is below 1\%. The MLP response is divided into eight intervals with boundaries of 0.0, 0.25, 0.4, 0.5, 0.6, 0.7, 0.8, 0.9 and 1.0. The MLP response is required to be greater than 0.25 in the final $p\muon^{-}$ mass distribution fit.

\begin{figure}
\centering
\includegraphics[scale=0.35]{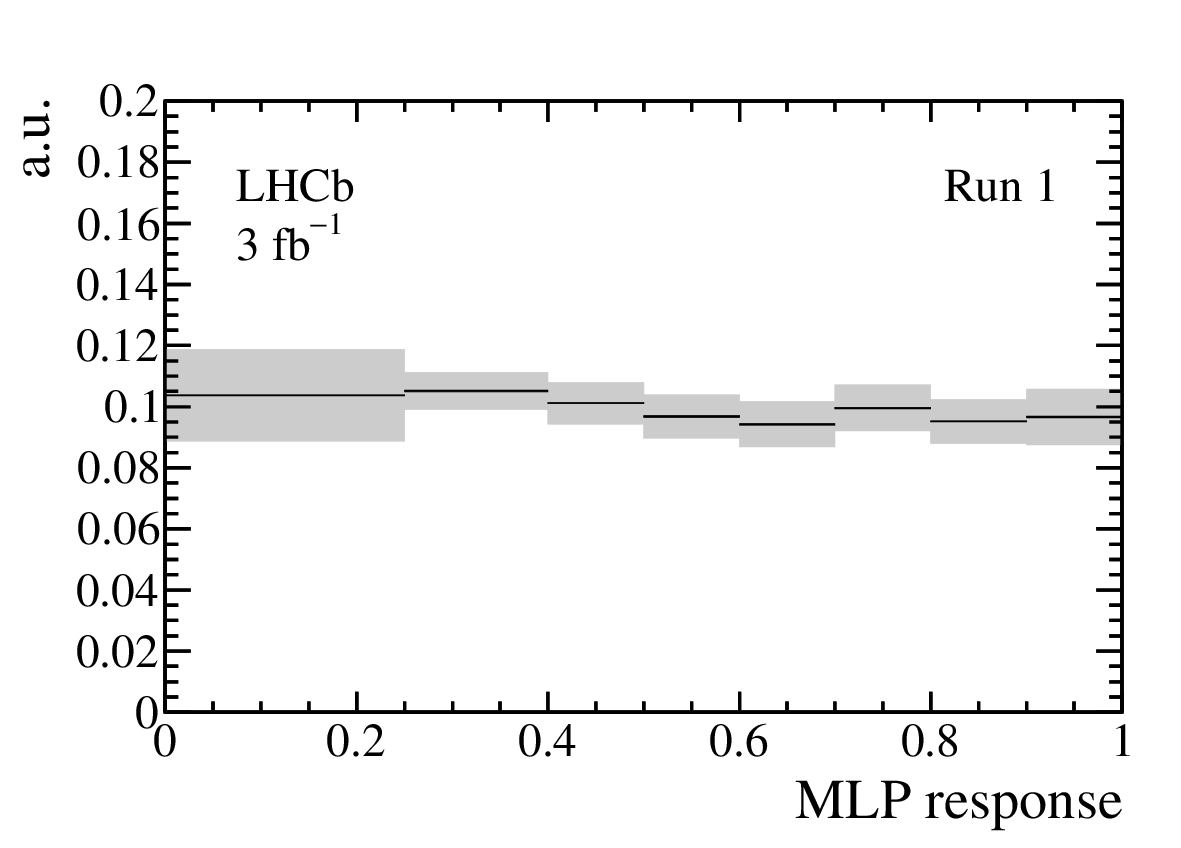}\includegraphics[scale=0.35]{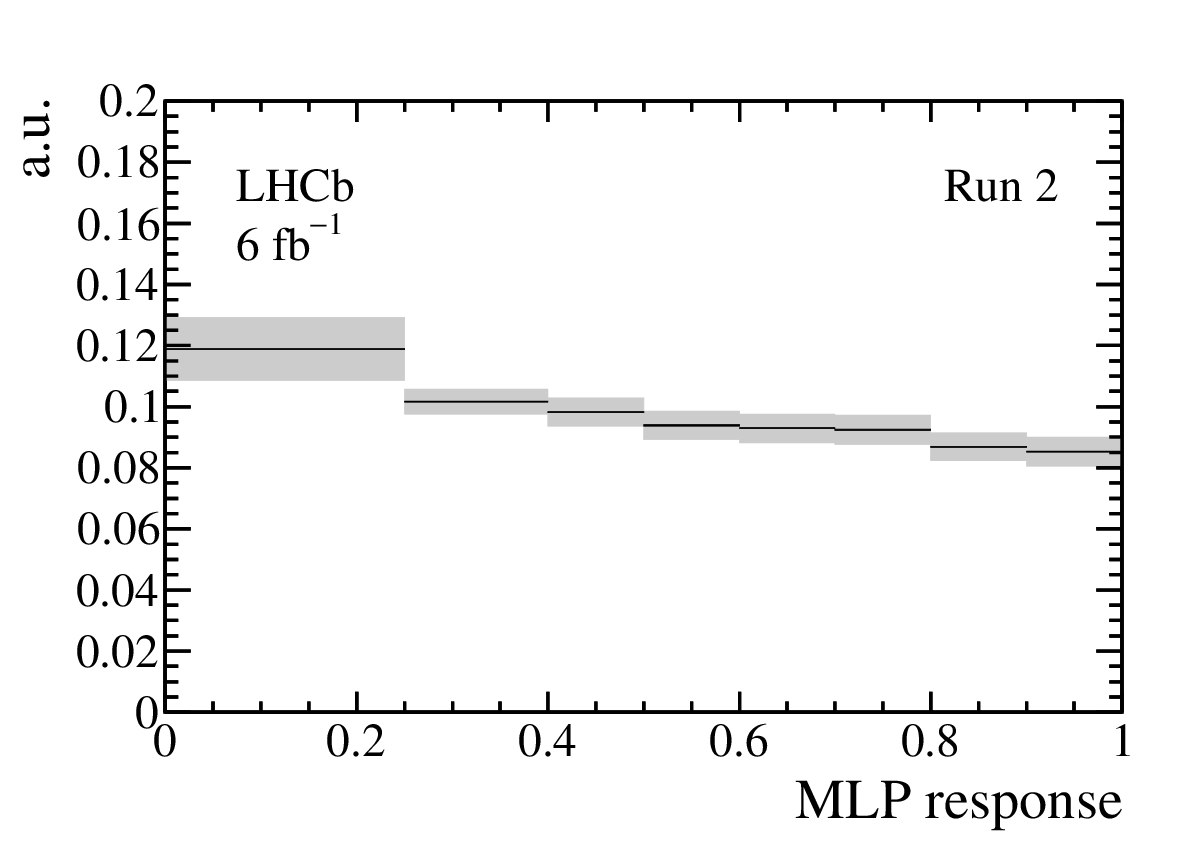}
\caption{Expected distribution of the MLP response for \BdsTopmu decays as obtained from the \BdToKpi control channel, for (left) Run 1 and (right) Run 2 data. The total (statistical and systematic combined) uncertainty is shown as a light grey band, and the thickness of each horizontal line at the centre indicates the statistical uncertainty. Each interval is normalised to its width.}
\label{fig:MLPPDF}
\end{figure}

Since the MLP classifier is trained using only kinematic information of a two-body \Bd decay, its response is calibrated using \BdToKpi decays. To avoid biases, \BdToKpi candidates are selected from candidates where the trigger decision does not depend on the \BdToKpi candidates themselves. The number of \BdToKpi candidates in each interval of the MLP response is determined by fitting the $K^{+}\pi^{-}$ invariant mass distribution after the full selection. Furthermore, the candidates are corrected for PID efficiency and weighted to emulate the effect of the \BdsTopmu triggers. The distributions of the MLP response on \BdToKpi simulated samples and data samples show good agreement. The distribution of the MLP response from the \BdToKpi data samples is corrected by the ratio of the \BdTopmu and \BdToKpi MLP responses in simulated samples. The corrected MLP response is shown in Fig.~\ref{fig:MLPPDF} and regarded as the expected probability density function (PDF) of the MLP response for \BdsTopmu decays, due to negligible kinematic differences between the $B^0$ and $B^0_s$ decays.

\section{Normalisation}
The \BdsTopmu yields are obtained from a fit to the $p\muon^{-}$ invariant mass distribution and translated into branching fractions with the normalisation channel \BuToJPsimuK, according to
\begin{eqnarray}
\BR(\BdsTopmu)=\frac{f_{\textup{norm}}}{f_{\textup{sig}}}\frac{\epsilon_{\textup{norm}}}{\epsilon_{\text{sig}}}\frac{N_{\text{sig}}}{N_{\textup{norm}}}\BR_{\textup{norm}} \equiv \alpha_{\text{sig}}N_{\text{sig}},
\label{con:norm}
\end{eqnarray}
where \BR, $\epsilon$, and $N$ are the branching fraction, efficiency and yield of the corresponding channel and $f_{\text{sig(norm)}}$ indicates the fragmentation fraction of the relevant $\B$ meson. The parameter $\alpha_{\text{sig}}$ is the single-event sensitivity. 

\begin{figure}
\centering
\includegraphics[scale=0.4]{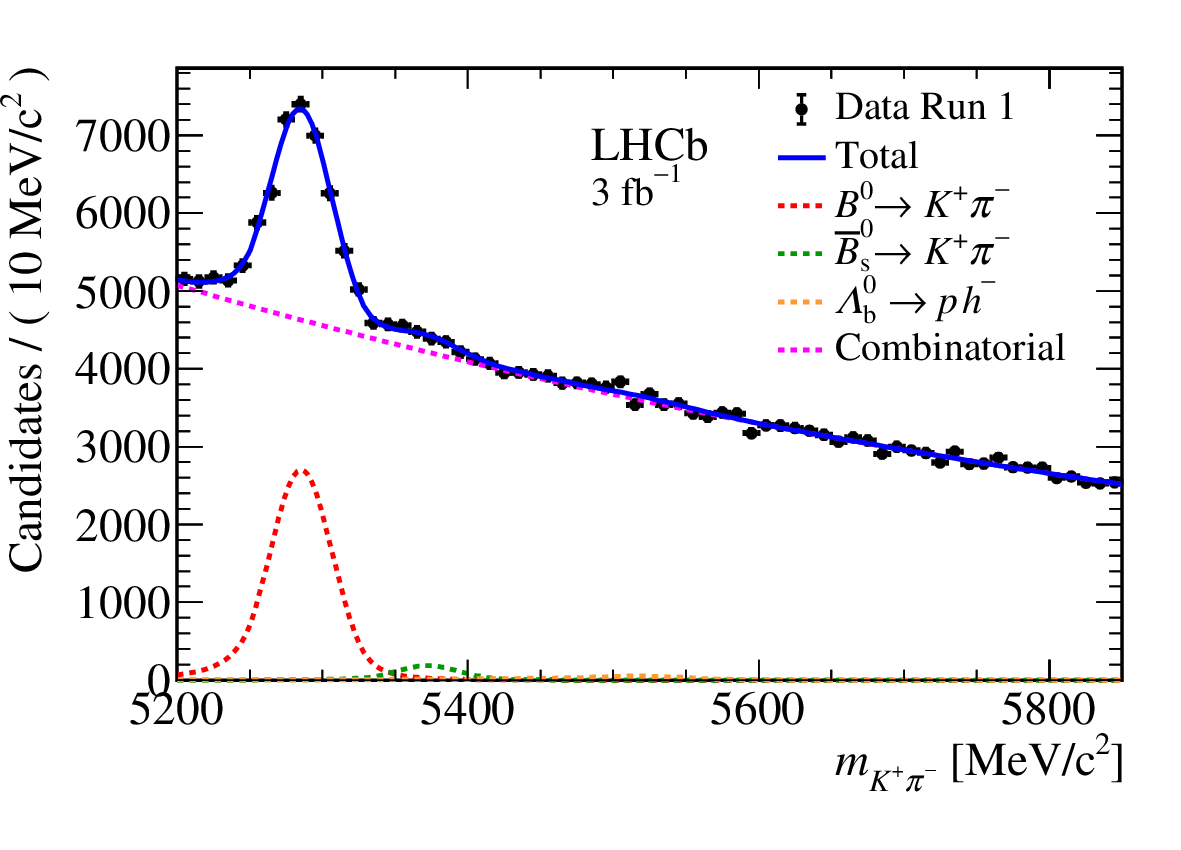}\includegraphics[scale=0.4]{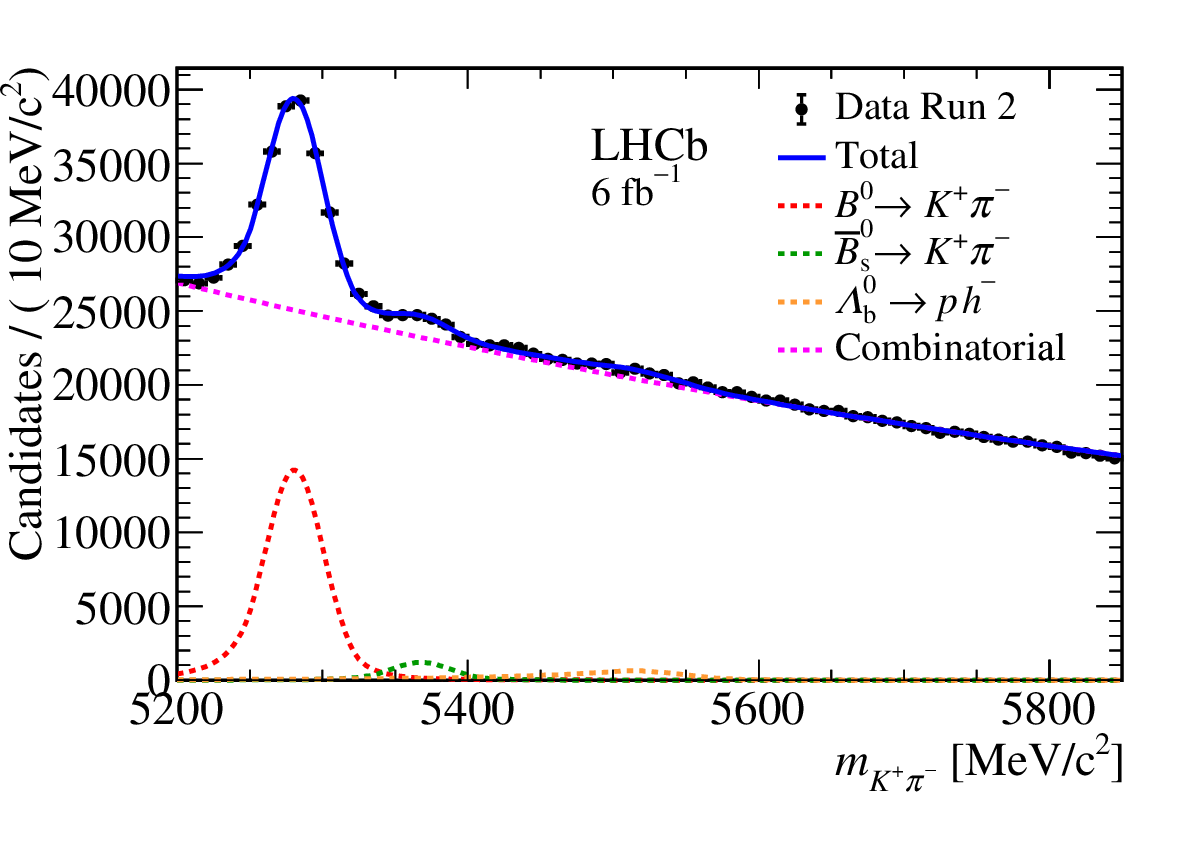}
\caption{ Mass distribution of \BdToKpi candidates in data for (left) Run 1 and (right) Run 2. The results of the invariant mass fits are superimposed.}
\label{fig:BToKpifit}
\end{figure}

\begin{figure}
\centering
\includegraphics[scale=0.4]{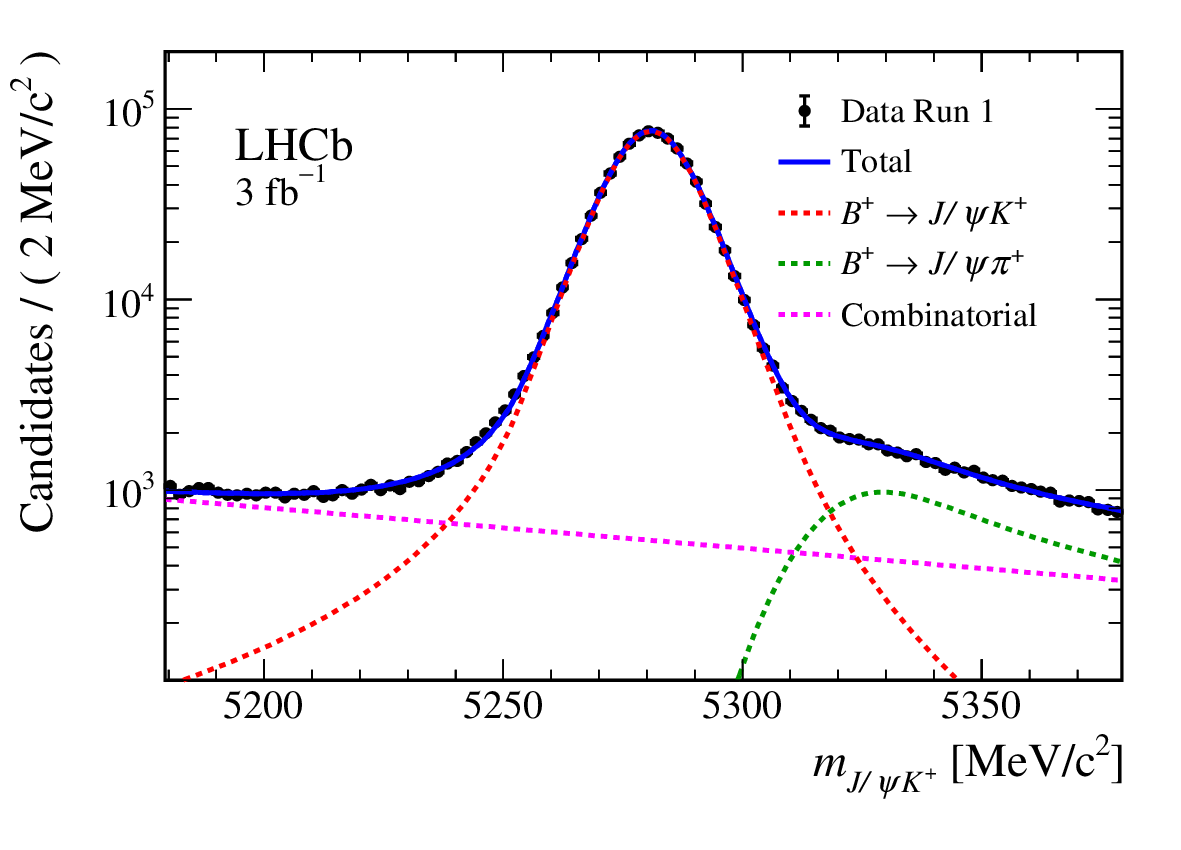}\includegraphics[scale=0.4]{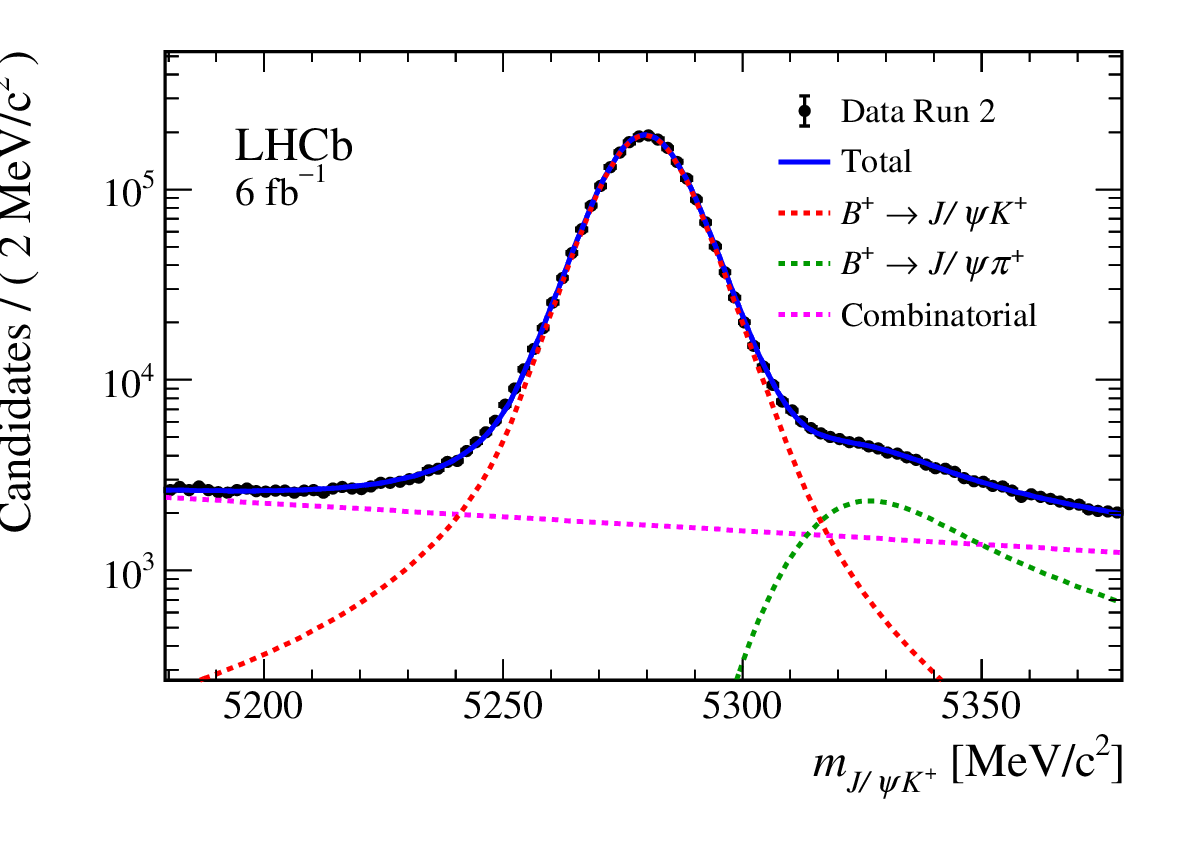}
\caption{ Mass distribution of \BuToJPsiK candidates in data for (left) Run 1 and (right) Run 2. The results of the invariant mass fits are superimposed.}
\label{fig:BToJPsiKfit}
\end{figure}

The \BdToKpi and \BuToJPsimuK decays are used as control channels. The \BdToKpi mode is used as a proxy to determine the MLP PDF for signal channels (Sec.~\ref{subsec:MLPCalibration}) and the \BuToJPsimuK mode is used to extract the signal branching fraction. To validate the normalisation procedure, the ratio between the measured branching fractions of \BdToKpi and \BuToJPsimuK is determined as
\begin{eqnarray}
R_{\textup{norm}}\equiv\frac{\BR \left(\BdToKpi\right)}{\BR (\BuToJPsimuK)}=\frac{N_{\BdToKpi}\times\varepsilon_{\BuToJPsimuK}}{N_{\BuToJPsimuK}\times\varepsilon_{\BdToKpi}},
\label{con:crosscheck}
\end{eqnarray}
where $\varepsilon$ and $N$ are the selection efficiency and yield. Using the same fit models as in Ref.~\cite{LHCb-PAPER-2021-008}, the yields of the normalisation channels are obtained through fits to the mass distributions of the candidates separately for Run 1 and Run 2 datasets. The results of the fits are shown in Figs.~\ref{fig:BToKpifit} and~\ref{fig:BToJPsiKfit}.

The selection efficiency for signal and normalisation channels includes efficiencies due to detector acceptance, tracking, reconstruction, trigger, and PID. All parts of the efficiency are evaluated using simulation, in which trigger and PID efficiencies are corrected using data~\cite{LHCb-PUB-2014-039,LHCb-PUB-2016-021}.

Calibration samples where the trigger decision is independent of the candidate decay products are used to study the trigger efficiency. From these samples, \BuToJPsimuK candidates are used to estimate the trigger efficiency for muons as a function of the muon \pt and IP. No proton trigger is used in this analysis. For the two normalisation channels, \BuToJPsimuK data samples are also used to determine the trigger efficiency map as a function of the $B$ meson \pt. The resulting trigger efficiency maps are then applied to the simulation to determine the integrated trigger efficiency for a specific channel. 

Particle identification efficiencies are evaluated using high-purity control samples of each particle species obtained from data~\cite{LHCb-PUB-2016-021,Aaij:2018vrk}. These control samples are obtained by means of kinematic requirements only, with muons obtained from \JpsiTomumu and \BuToJPsiK decays, pions and kaons from \DToKpi decays selected via $D^{*+}\rightarrow \Dz\pip$ decays, and protons from $\Lz\rightarrow \proton \pim$ and $\Lz_c^{+}\rightarrow\proton\Km\pip$ decays. The muon PID efficiencies are evaluated as a function of the muon \pt and $\eta$. The PID efficiencies for different hadrons are evaluated as a function of \ptot and $\eta$. The single-track efficiencies are then combined and averaged using the kinematic distributions of the corresponding simulated sample. 

The ratios $R_{\textup{norm}}$ between the measured branching fractions of \BdToKpi and \BuToJPsimuK in Eq.~\ref{con:crosscheck} are $R_{\textup{norm}}$ = 0.329 $\pm$ 0.012 and $R_{\textup{norm}}$ = 0.341 $\pm$ 0.010 for Run 1 and Run 2, respectively, where the uncertainties are statistical only and include the efficiency uncertainties from both channels. These values are in agreement with the ratio of the world averages of these branching fractions, $R_{\textup{PDG}}$ = 0.326 $\pm$ 0.012~\cite{ParticleDataGroup:2020ssz}.

\section{Background contributions}
In addition to the combinatorial background, the signal region is also potentially contaminated by background contributions from exclusive decays where one or more of the final-state particles are misidentified or not reconstructed. The most dangerous of these backgrounds
are hadronic $X_b\rightarrow h^{+}h^{\prime-}$ decays, such as: \BdToKpi, \BsToKK, \BsToKpi, \LbTopK and \LbToppi, and partially reconstructed semileptonic decays: \LbTopmunu, \BsToKmunu and \BdTopimunu. The partially reconstructed decays do not peak in the signal region, but are potentially abundant. The expected number of candidates from each possible background decay that pass the signal selection is evaluated using simulation. The candidates from $B_{(s)}^0$ decays are normalised to the number of \BuToJPsimuK decays found in data as
\begin{eqnarray}
N_{X}=N_{\BuToJPsimuK}\cdot\frac{f_{q}}{f_{u}}\cdot\frac{\BR(X)}{\BR(\BuToJPsimuK)}\cdot\frac{\varepsilon(X)}{\varepsilon(\BuToJPsimuK)},
\end{eqnarray}
where $N_X$ is the expected number of candidates from the $X$ decay that fall into the \BdsTopmu signal mass window; $N_{\BuToJPsimuK}$ is the yield of \BuToJPsimuK decays in the data; $f_q$ is the fragmentation fraction for a \bquark quark to produce either a meson with secondary quark content $q$ or a baryon; 
and $\BR$ and $\epsilon$ are the branching fraction and selection efficiency for a particular channel, respectively. 
For \Lb decays, the fragmentation fraction $f_{\Lb}/f_d$ depends on the \pt of the \Lb baryon~\cite{LHCb:2014ofc,LHCb:2019fns}. The $\LbTopK$ decay is used as a normalisation channel to estimate the number of background $\Lb$ candidates, thereby removing a potential bias due to the correlation between $f_{\Lb}/f_d$ and \pt. The selection criteria on $\LbTopK$ candidates are largely the same as for the \BdToKpi normalisation channel except for the $p$ and $K$  identification requirements.

The expected total number of $X_b\rightarrow h^{+}h^{\prime-}$ decays in the full MLP range is found to be negligible. The only background sources which are found to be relevant are the semileptonic decays \LbTopmunu, \BsToKmunu and \BdTopimunu, with yields estimated to be roughly $8500 \pm 2600$, $50\pm 13$ and \mbox{$310 \pm 16$}, respectively, for Run 1 and $21000 \pm 6400$, $64 \pm 17$ and $410 \pm 23$, respectively, for Run 2. The Run 2 to Run 1 yield ratios of \BsToKmunu and \BdTopimunu are considerably smaller than that of \LbTopmunu, which is mainly due to improved discrimination of protons from pions and kaons in Run 2. These three decay modes are included in the final fit model.


\section{Measurement of signal branching fractions}
The branching fractions of the signal decays are determined using an unbinned extended maximum-likelihood fit to the $p\muon^{-}$ mass distributions, performed simultaneously on all the subsets. 
The fit is performed in the mass window
$5067 < m_{p\muon^{-}} < 5667$ \mevcc. The full fit model in the 
data-taking period $r$ and 
MLP interval $i$, ${\rm PDF}_{\rm Full}^{r,i}(m)$ is:
\begin{small}
\begin{align}
{\rm PDF}_{\rm Full}^{r,i}(m) &=\mathcal{P}\left(N_{\rm meas}|{N_{{\rm com},r,i}}+\sum_jN_{r,j}F_{r,i,j}+N_{{\rm sig},r,i}\right)\times \frac{1}{{N_{{\rm com},r,i}}+\sum_jN_{r,j}F_{r,i,j}+N_{{\rm sig},r,i}}\nonumber\\
&\times \bigg({N_{{\rm com},r,i}}{\rm PDF}_{{\rm com},r}(m)+\sum_jN_{r,j}F_{r,i,j}{\rm PDF}_{r,i,j}(m)+N_{{\rm sig},r,i}{\rm PDF}_{\rm sig}(m)\bigg),
\label{con:sigequation1}
\end{align}
\end{small}
where $\mathcal{P}$ is the Poisson probability of measuring $N$, given the fitted yields. The signal model, denoted ${\rm PDF}_{\rm sig}$, is described in detail below. Also $\rm{PDF_{com,{\it r}}}$ is the PDF for combinatorial background, which is described with an exponential function with an independent shape parameter in each data-taking period and independent yield $N_{{\rm com},r,i}$ in each subset. $\rm{PDF_{\it r,i,j}}$ represents the PDF for 
the physical background with index $j$, which is modelled based on histograms determined from simulation and smoothed with a second order interpolation. $N_{r,j}$ and $F_{r,i,j}$ represent the expected yield and the yield fraction in each subset for physical backgrounds. 

${\rm PDF}_{\rm sig}$ is a sum of a double-sided Crystal Ball (DSCB) function~\cite{Skwarnicki:1986xj} and a Gaussian function. 
The signal shape parameters are obtained from simulation, with data-driven scale factors applied to the core resolution to correct for minor data-simulation discrepancies. For this purpose, since there is no appropriate control channel with a proton and a muon in the final state, \LbTopK and \JpsiTomumu decays are analysed comparing the invariant mass resolution in data and simulation. The results are then combined to reproduce the effect on
an $p\muon^{-}$ final state. Corrections to the mass resolution are of the order of 10\%.  The mass shape parameters are found to be independent of the particular MLP interval and only two models, for the Run~1 and Run~2 data samples, are used.

The yield in each subset, $N_{{\rm sig},r,i}$, is expressed as

\begin{align}
N_{{\rm sig},r,i} &=\frac{\BR_{\rm sig}{f_{\rm sig}}\varepsilon_{{\rm PID},r,i}\varepsilon_{{\rm others},r}F_{{\rm sig},r,i}{N_{{\rm norm},r}}}{{\BR_{\rm norm}f_{\rm norm}\varepsilon_{{\rm norm},r}}}.
\label{con:sigequation2}
\end{align}
Here, $\BR$ and $f$ represent the branching fraction and fragmentation fraction of the signal channel or the normalisation channel. The quantities $N_{{\rm norm},r}$ and $\varepsilon_{{\rm norm},r}$ are the yield and efficiency of the normalisation mode \BuToJPsimuK. Furthermore, $\varepsilon_{{\rm PID},r,i}$ and $\varepsilon_{{\rm others},r}$ are the PID efficiency and the total efficiency without the PID requirements for signal channels.
$F_{r,i,j}$ represents the fraction in each subset for signal channels (Fig.~\ref{fig:MLPPDF}). The quantities $\BR_{\rm sig}$ and $N_{{\rm com},r,i}$ are free to vary in the fit, while $f_{\rm sig}$, $N_{{\rm norm},r}$, $\BR_{\rm norm}$, $f_{\rm norm}$, $\varepsilon_{{\rm norm},r}$, $N_{r,j}$, $F_{r,i,j}$, $\varepsilon_{{\rm PID},r,i}$, $\varepsilon_{{\rm others},r}$, $F_{{\rm sig},r,i}$ and $N_{{\rm norm},r}$ are Gaussian-constrained, according to the expected value and uncertainty.

\begin{figure}
\centering
\includegraphics[scale=0.37]{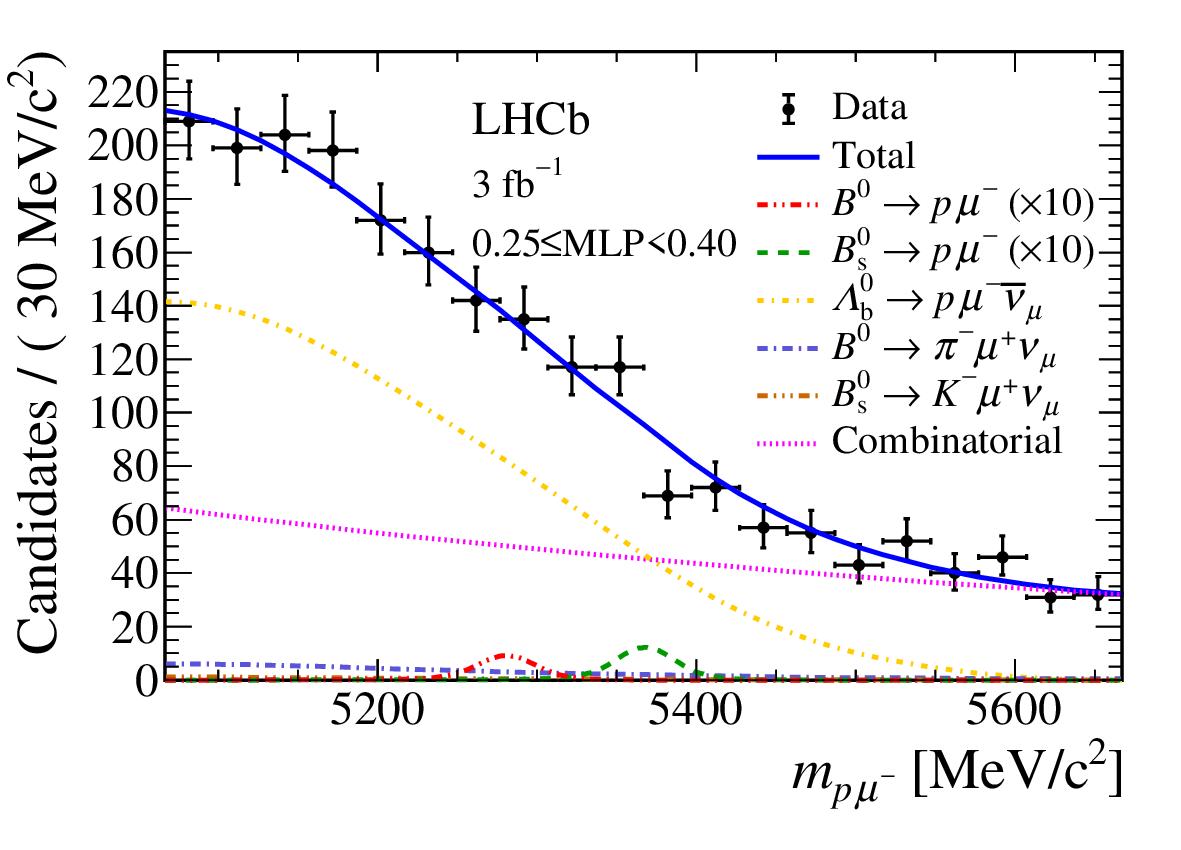}\includegraphics[scale=0.37]{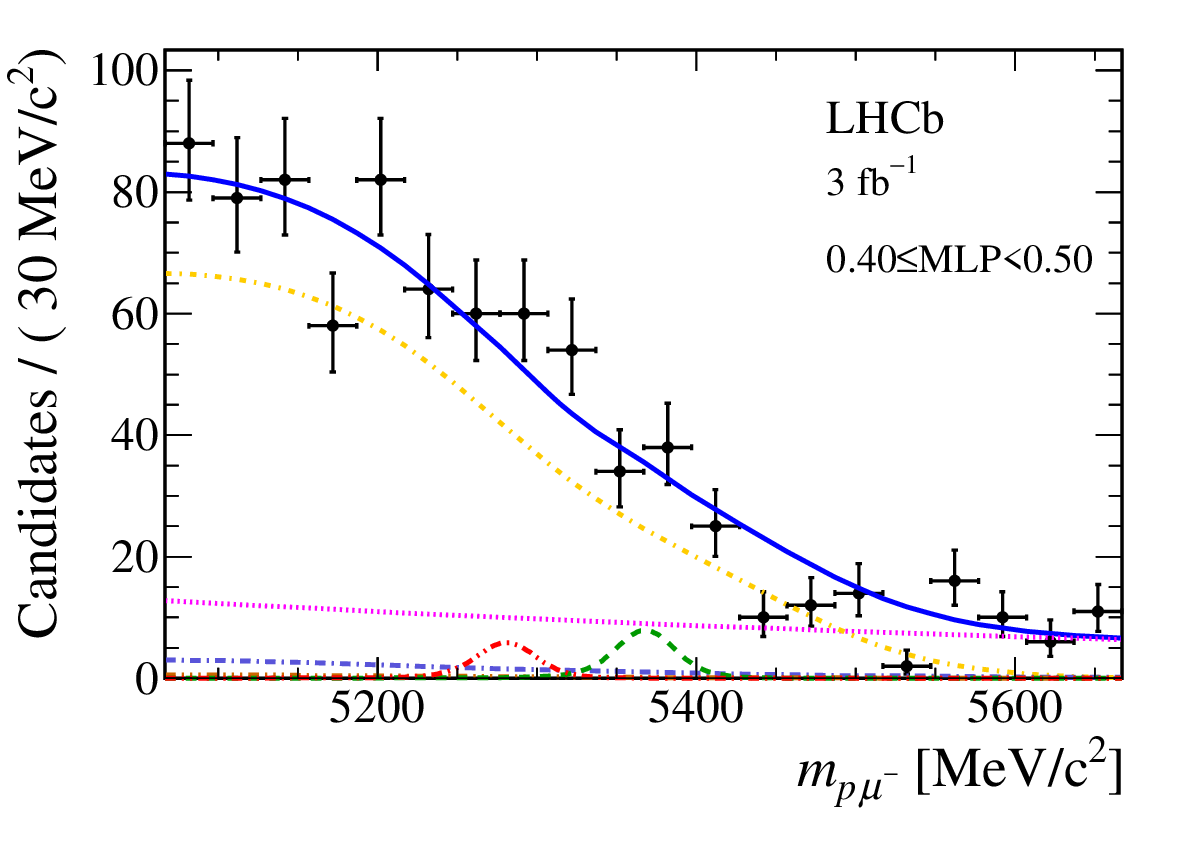}
\includegraphics[scale=0.37]{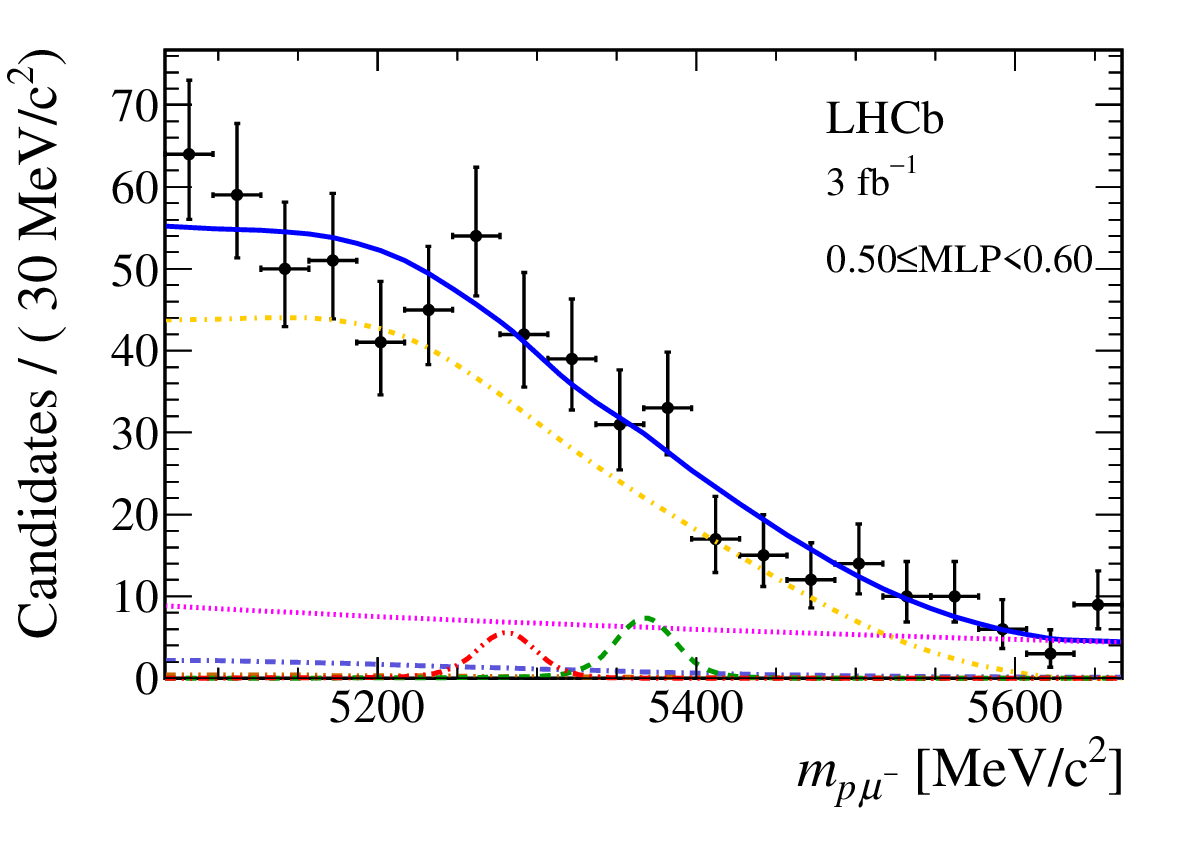}\includegraphics[scale=0.37]{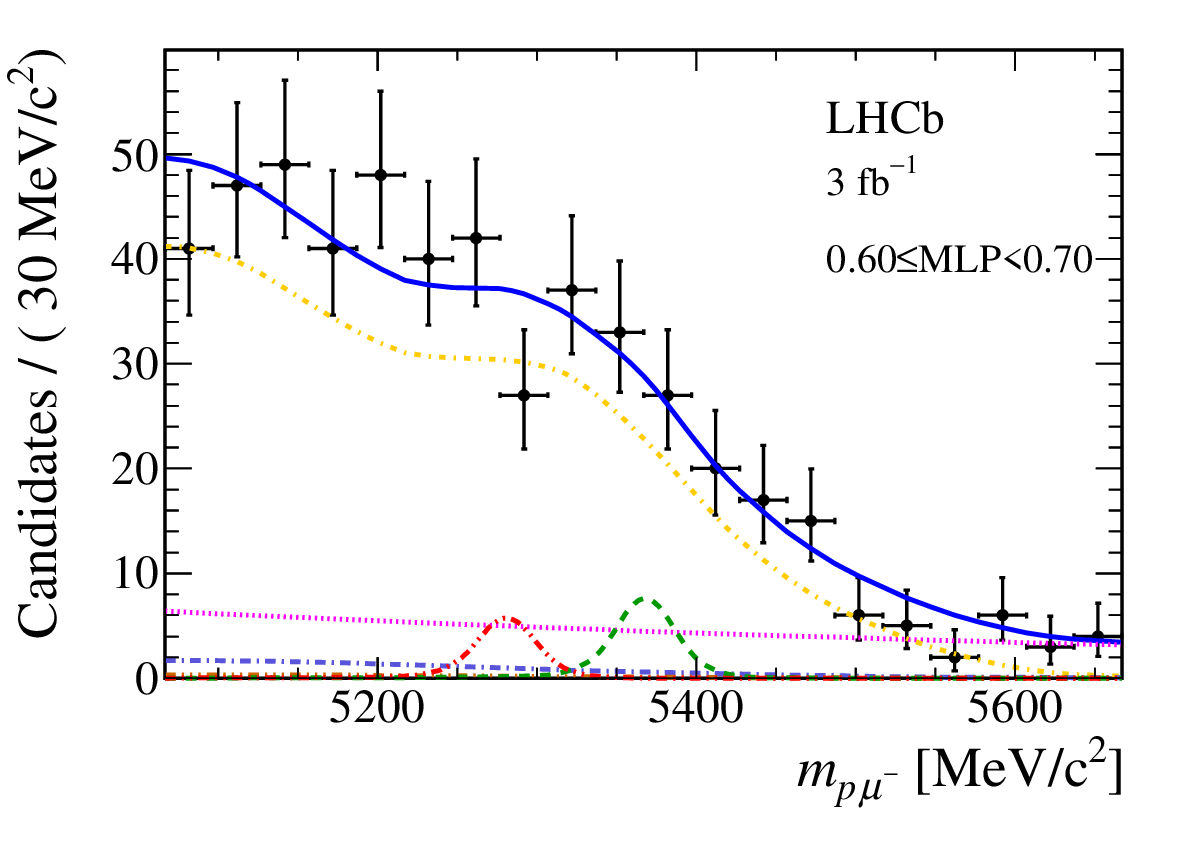}
\includegraphics[scale=0.37]{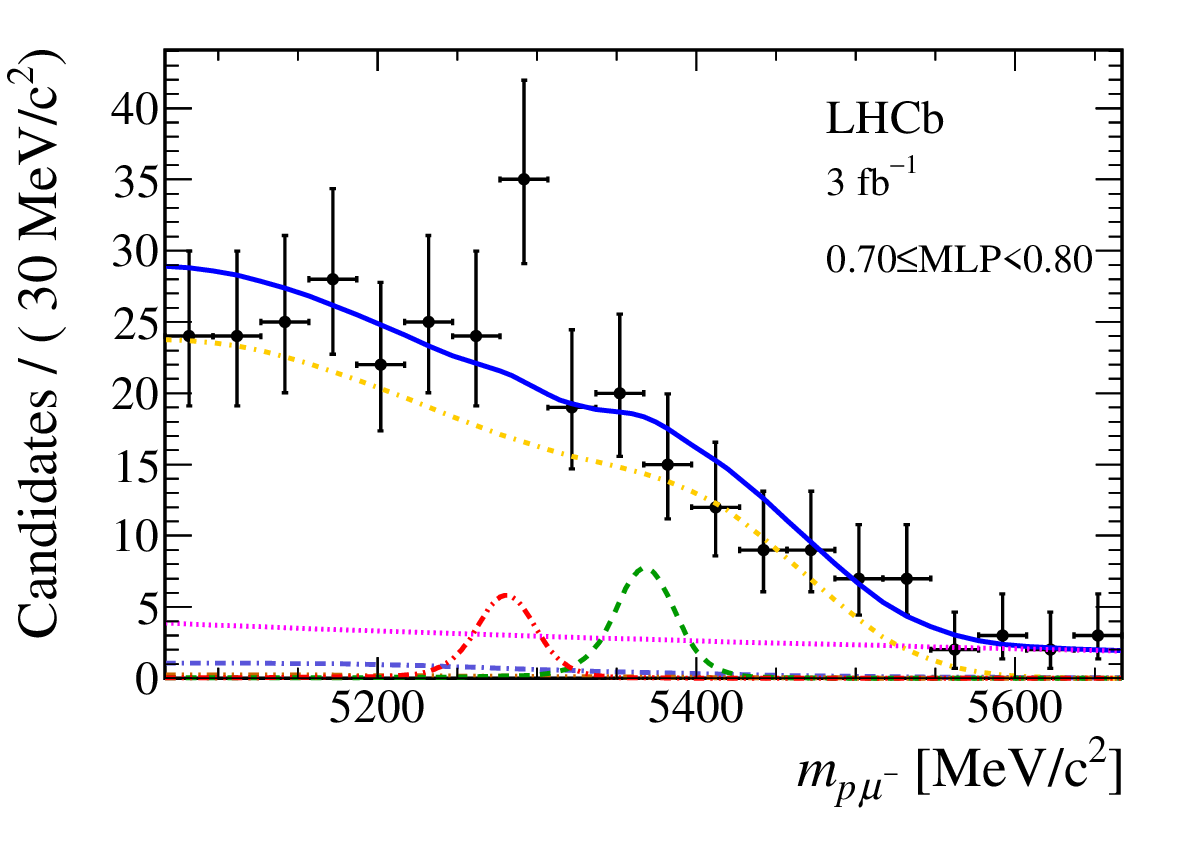}\includegraphics[scale=0.37]{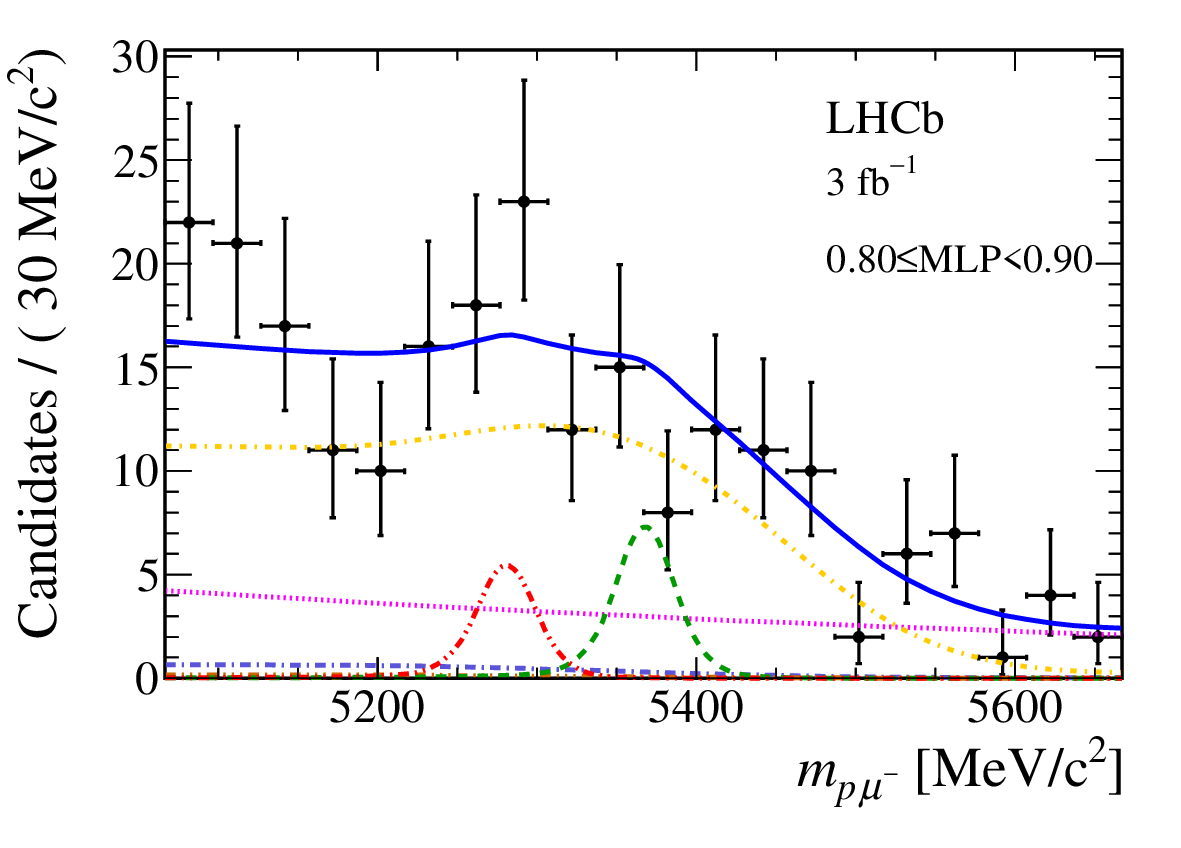}
\includegraphics[scale=0.37]{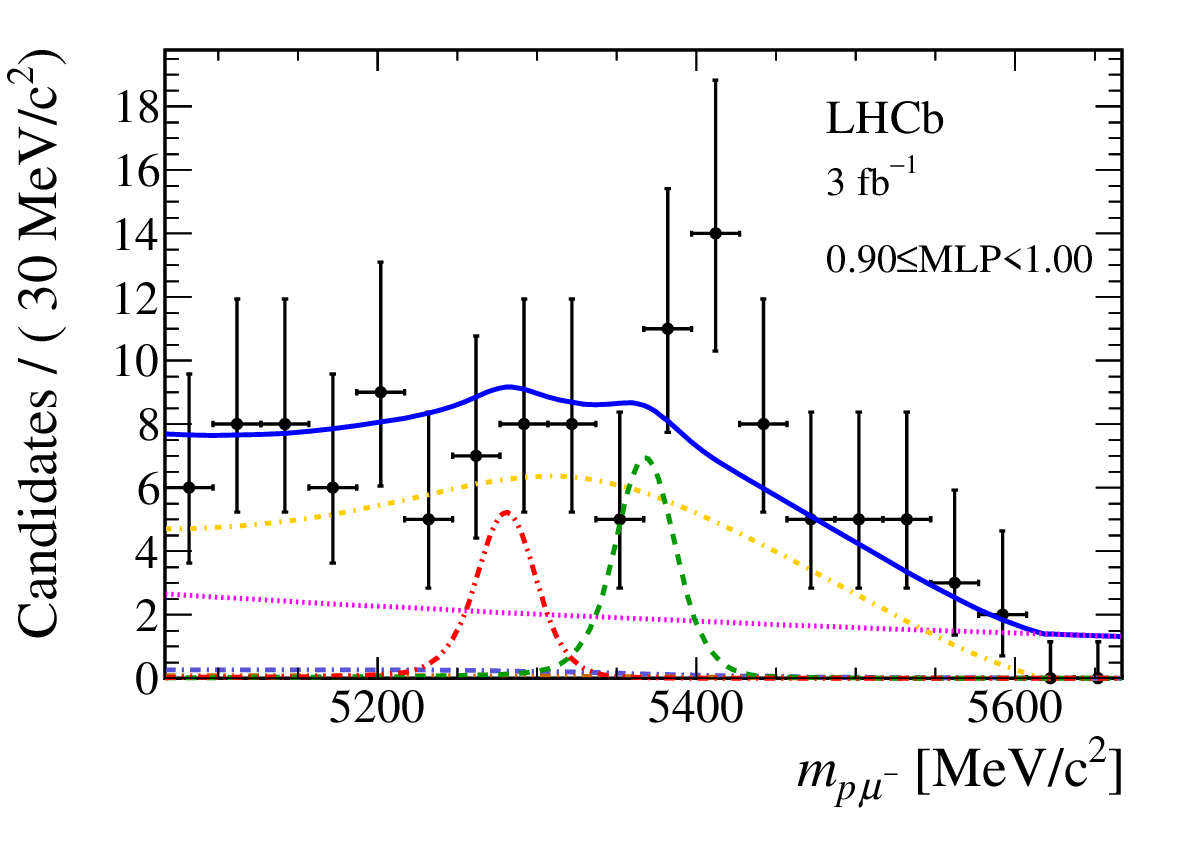}
\caption{Mass distribution of signal candidates (black dots) for Run 1 samples in regions of MLP. The results of the fit described in the text are also shown. The contributions from the two signal channels, \BdTopmu(red dashed lines) and \BsTopmu(green dashed lines), are scaled by a factor of 10 for better visibility.}
\label{fig:finalfit1}
\end{figure}

\begin{figure}
\centering
\includegraphics[scale=0.37]{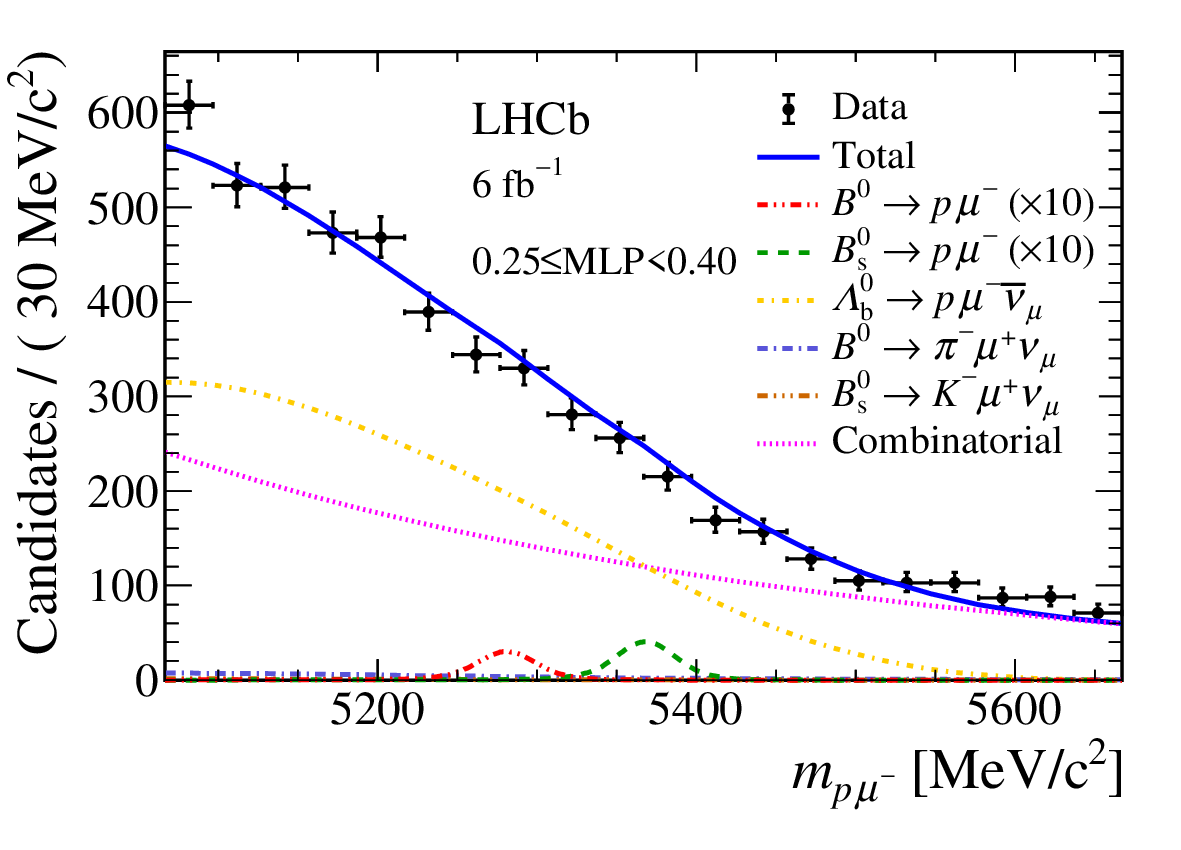}\includegraphics[scale=0.37]{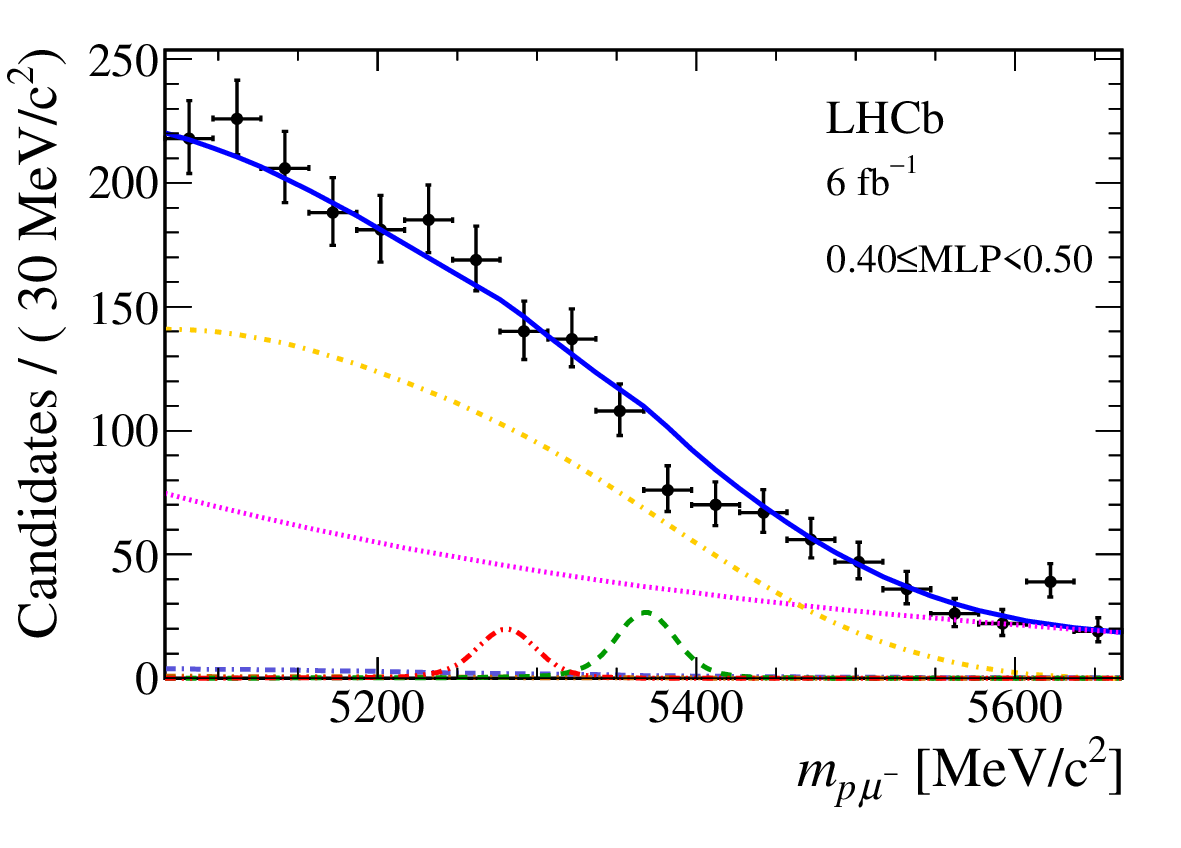}
\includegraphics[scale=0.37]{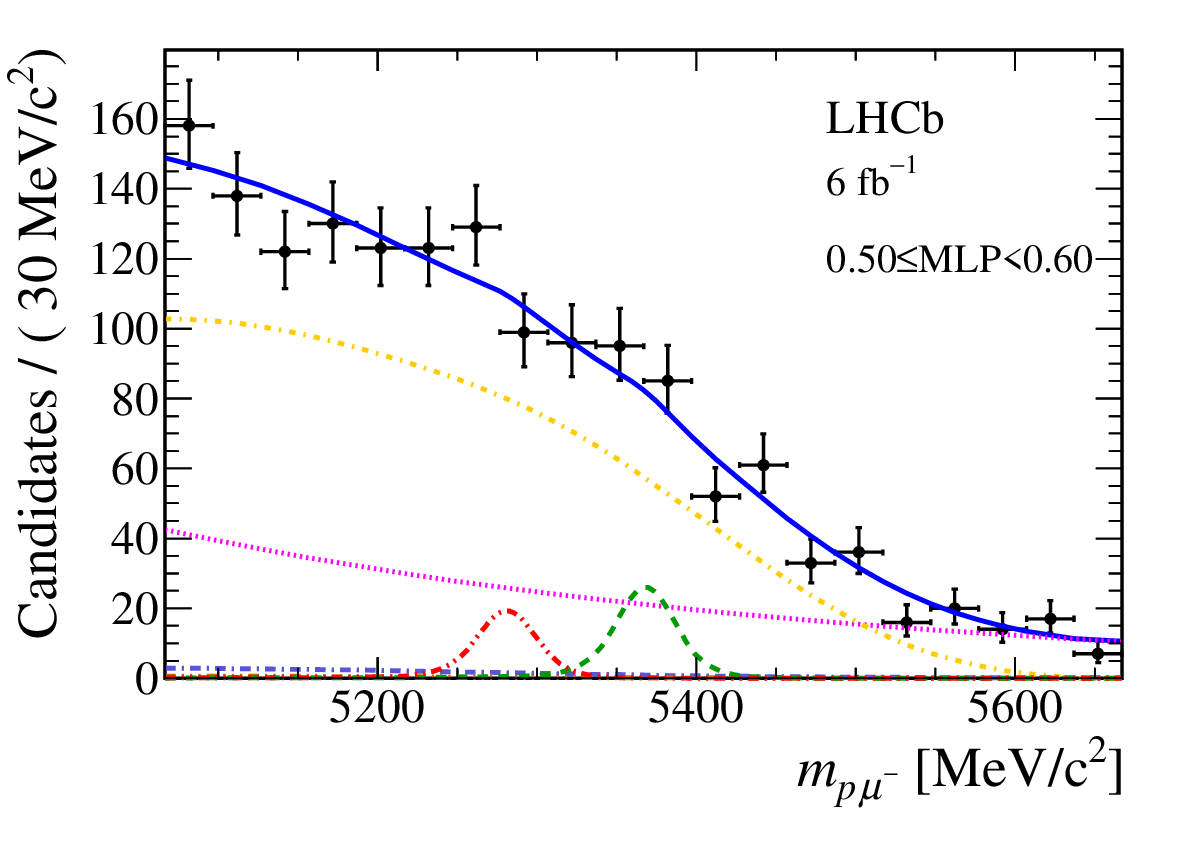}\includegraphics[scale=0.37]{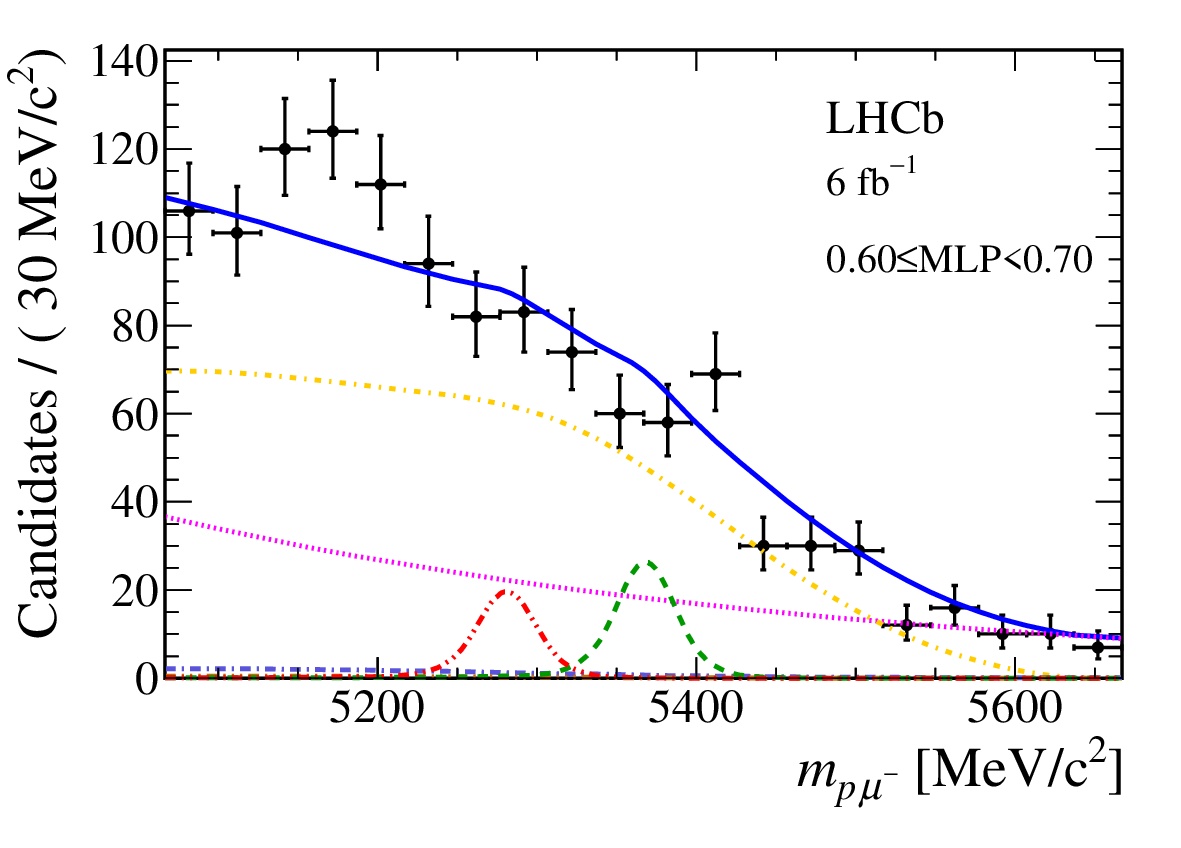}
\includegraphics[scale=0.37]{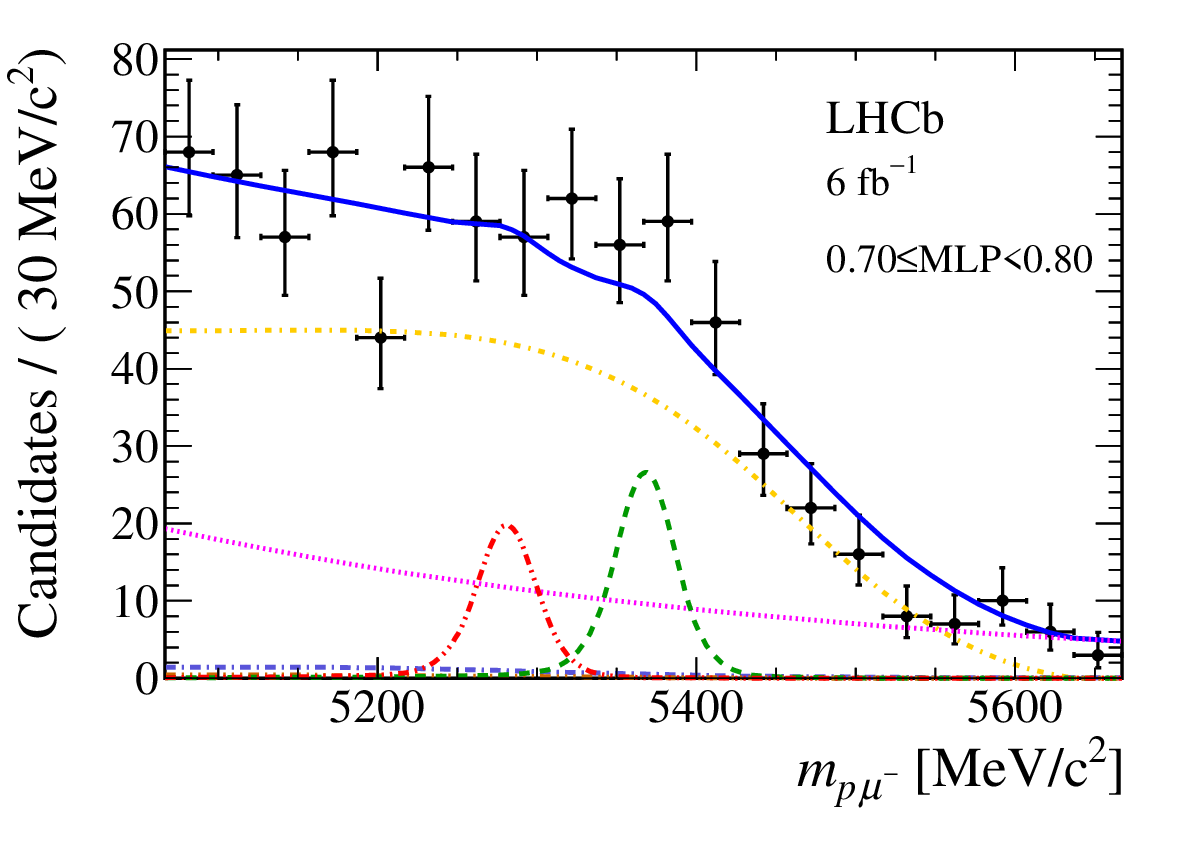}\includegraphics[scale=0.37]{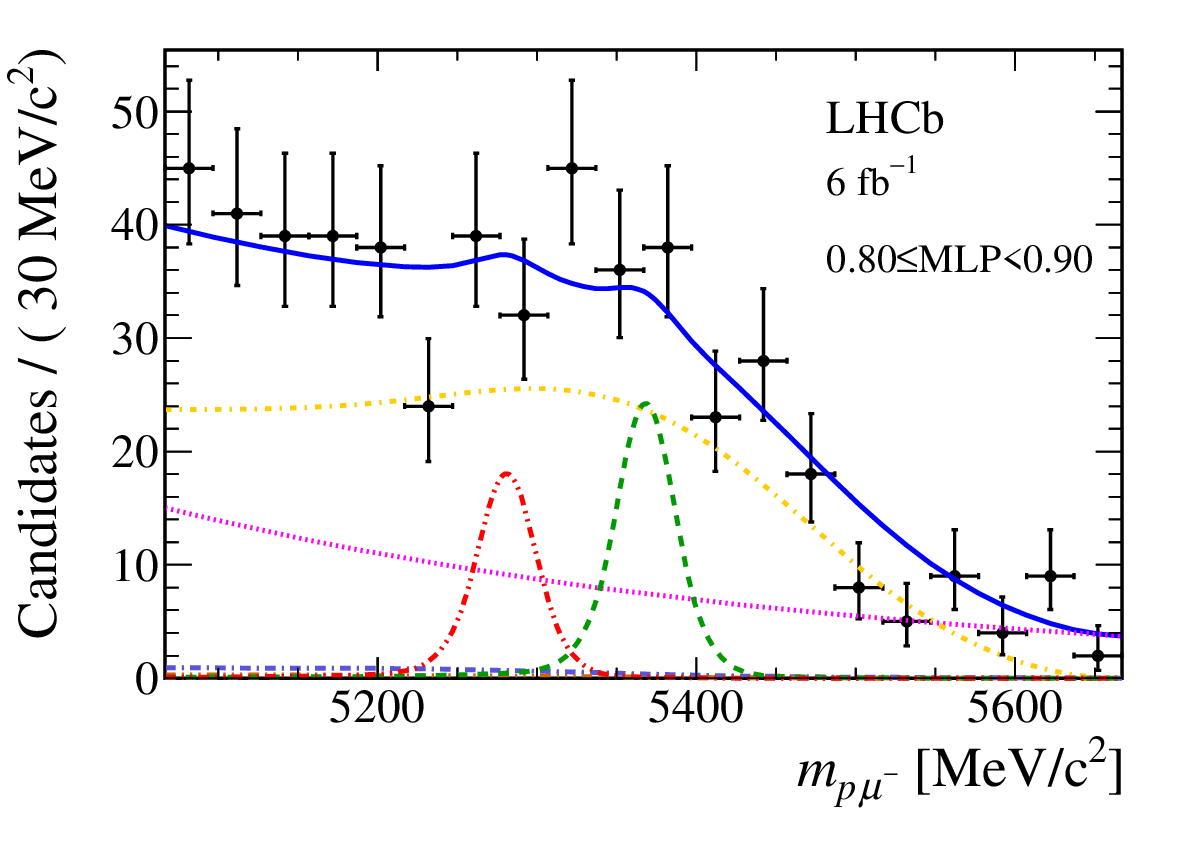}
\includegraphics[scale=0.37]{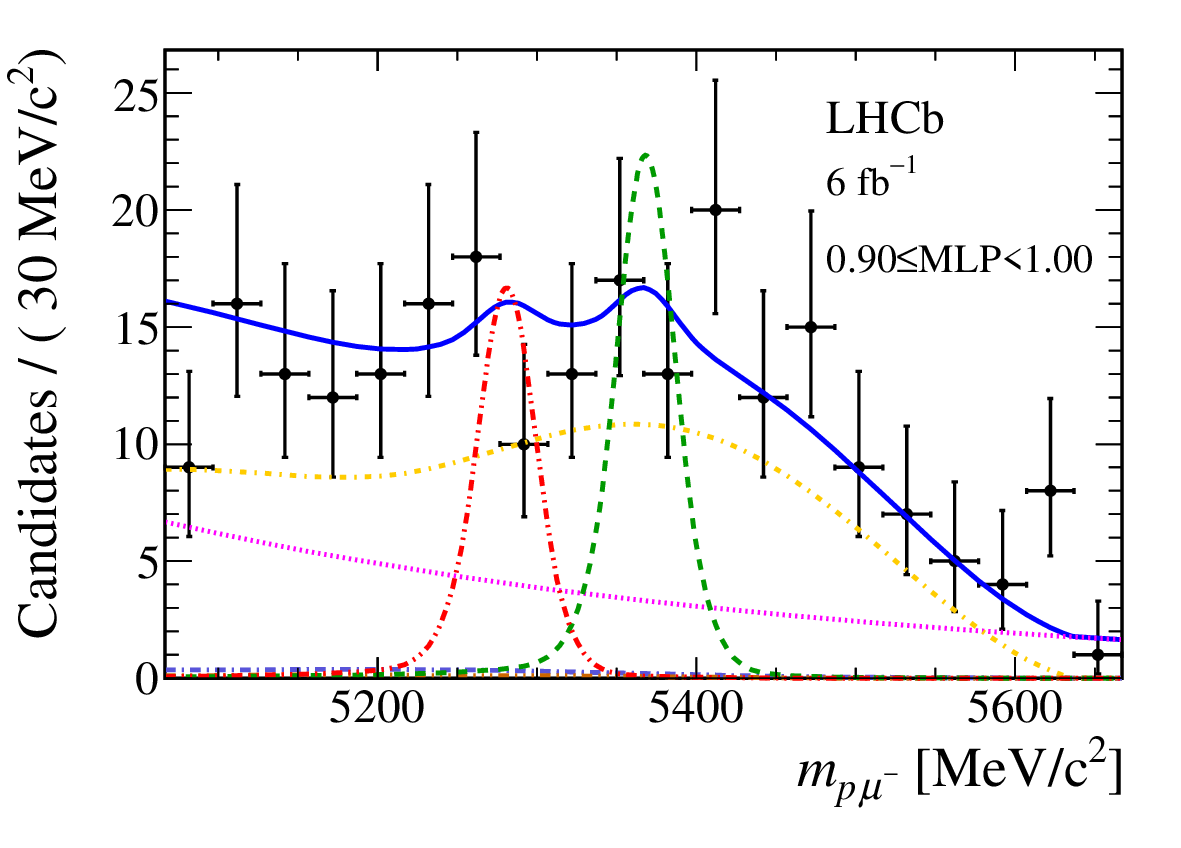}
\caption{Mass distribution of signal candidates (black dots) for Run 2 samples in regions of MLP. The results of the fit described in the text are also shown. The contributions from the two signal channels, \BdTopmu(red dashed lines) and \BsTopmu(green dashed lines), are scaled by a factor of 10 for better visibility.}
\label{fig:finalfit2}
\end{figure}

\begin{figure}
\centering
\includegraphics[scale=0.39]{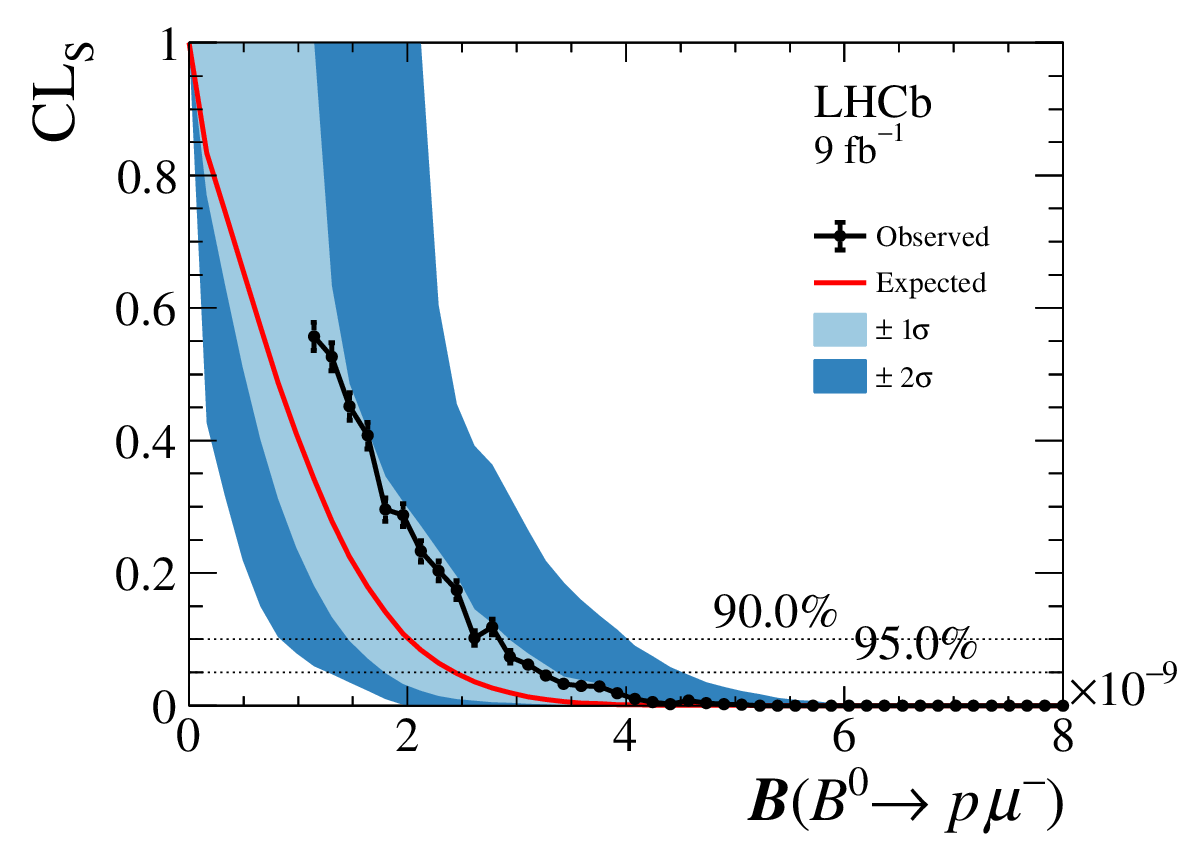}\includegraphics[scale=0.39]{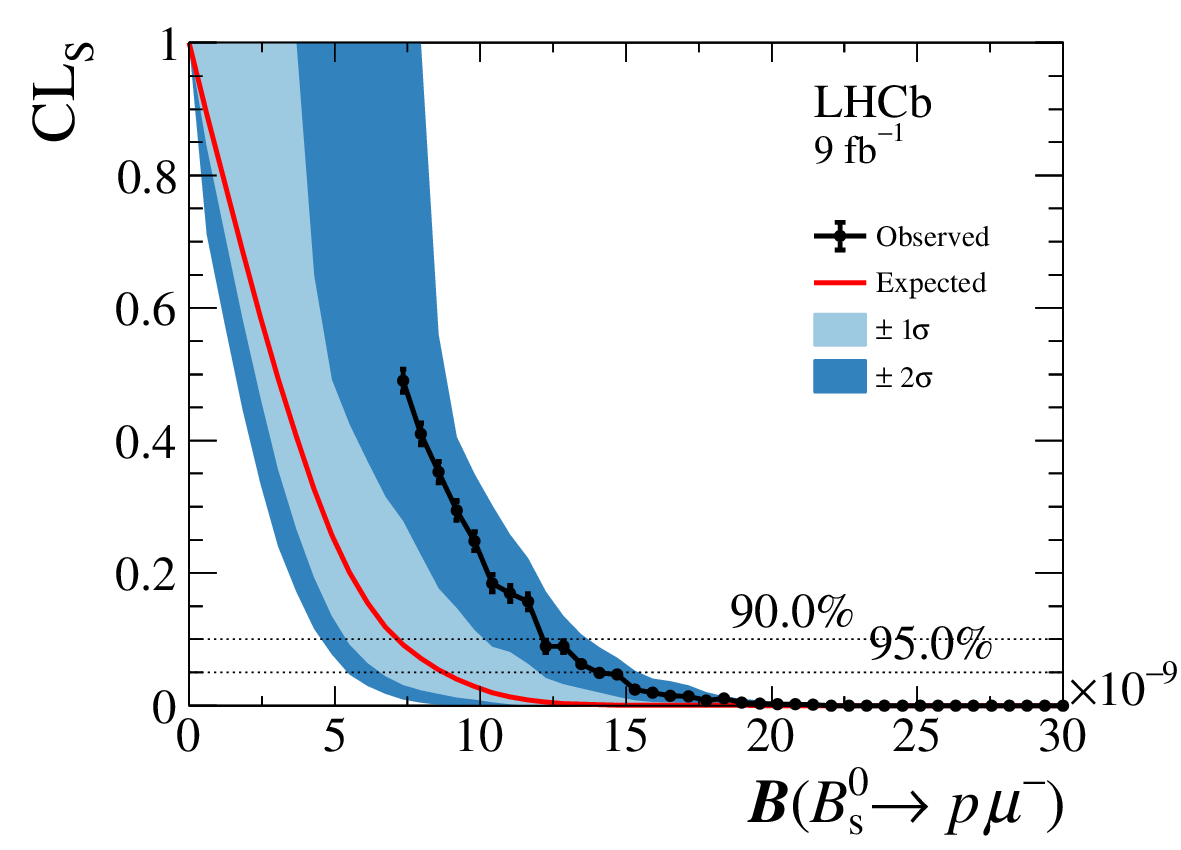}
\caption{ Results from the CLs scan used to obtain the limit on (left) $\BR(\BdTopmu)$ and (right) $\BR(\BsTopmu)$. The background-only expectation is shown by the red line and the 1$\sigma$ and 2$\sigma$ bands are shown as light blue and blue bands respectively. The observation is shown as the solid black line. The two dashed lines intersecting with the observation indicate the limits at
90\% and 95\% CL for the upper and lower line, respectively.}
\label{fig:ULdetermine}
\end{figure}

The result of the fit in each subset is shown in Figs.~\ref{fig:finalfit1} and~\ref{fig:finalfit2}. The resulting branching fractions 
are:
\begin{align}
\BR(\BdTopmu)=(0.84\pm1.17\pm0.57)\times 10^{-9}, \nonumber \\
\BR(\BsTopmu)=(4.28\pm3.99\pm2.29)\times 10^{-9}, \nonumber
\end{align}
where the first uncertainty is statistical and the second systematic. No significant excess of \BdTopmu or \BsTopmu decays is observed and upper
limits on the branching fractions are set using the ${\rm CL}_{\rm s}$ method~\cite{Read_2002} with a one-sided test statistic~\cite{Cowan:2010js} as implemented in Refs.~\cite{GammaCombo,LHCb:2016mag}. The one-sided test statistic for a given branching fraction value is defined as twice the negative logarithm of the profile likelihood ratio if it is larger than the measured branching fraction and zero otherwise. Its distribution is determined from pseudoexperiments, where nuisance parameters are set to their best fit values when generating pseudoexperiments. The central values of the Gaussian-constraints are
independently varied within their uncertainties for each pseudoexperiment as described in Ref.~\cite{LHCb:2009}. With the inclusion of systematic effects as discussed below, the ${\rm CL}_{\rm s}$ curves as a function of the branching ratios are shown in Fig.~\ref{fig:ULdetermine},  and the upper limits at \mbox{90 (95)\%} confidence level (CL) are reported in Table~\ref{tab:Expected-and-observed}, where the observed upper limits are both above the expected ones. This is likely due to a fluctuation of the concentration of $\LbTopmunu$ events around the known $B_{(s)}^0$ masses in MLP regions with large  MLP response values.

\begin{table}[h]
\caption{Expected and observed upper limits for $\BR(\BdTopmu)$ and $\BR(\BsTopmu)$ at 90\% (95\%) CL, with systematic effects included. No signal is assumed for the expected upper limits.}
\begin{center}
\begin{tabular}{ccc}
\toprule[1pt]
Channel & Expected &  Observed\tabularnewline
\hline
\BdTopmu & $1.9\ (2.4)\times10^{-9}$ & $2.6\ (3.1)\times10^{-9}$\tabularnewline
\BsTopmu & $7.0\ (8.6)\times10^{-9}$ & $12.1\ (14.0)\times10^{-9}$\tabularnewline
\toprule[1pt]
\end{tabular}
\end{center}
\label{tab:Expected-and-observed}
\end{table}

Several systematic uncertainties can affect the evaluation of the limit on the \BdTopmu and \BsTopmu branching fractions through the normalisation formula (Eq.~\ref{con:norm}). The systematic uncertainties considered are related to the fraction of signal and physical backgrounds in each MLP interval; expected total yields of physical backgrounds; signal efficiency; mass resolution of signal shape; fragmentation fraction $f_{\rm sig}/f_{\rm norm}$ (only applicable for the $B_s^0$ mode); branching fraction, efficiency and yield of the normalisation channel $\BuToJPsimuK$. The systematic uncertainties on the efficiencies of the signal and normalisation channels include the uncertainties 
arising from modelling the dependencies of the trigger (PID) efficiency maps. The effect on the $p\mu^-$ mass distribution and yield fraction from \LbTopmunu form factor uncertainty and different \LbTopmunu physics models have been studied and found to be negligible. No systematic uncertainty is assigned due to the assumptions of unpolarized final states. These systematic uncertainties are taken into account for the limit computation by constraining the respective parameters in the likelihood fit with a Gaussian distribution having the central value of the parameter as the mean and its uncertainty as the width. The overall
impact on the limits is evaluated to be 4\% on \BdTopmu and 11\% on \BsTopmu.


\section{Summary}
In summary, a search for the LNV and BNV decays \BdTopmu and \BsTopmu is performed
using the full Run 1 and Run 2 data samples of the \lhcb experiment, using a sample of proton-proton collision data corresponding to a total integrated luminosity of 9\invfb. No excesses are observed for these two modes and upper limits on the branching fractions are set to be $\BR(\BdTopmu) < 2.6\ (3.1)\times 10^{-9}$ and $\BR(\BsTopmu) < 12.1\ (14.0)\times 10^{-9}$ at 90\% (95\%) CL. These results represent the first upper limits on these decays to date.

\clearpage

\section*{Acknowledgements}
%
%
\noindent We express our gratitude to our colleagues in the CERN
accelerator departments for the excellent performance of the LHC. We
thank the technical and administrative staff at the LHCb
institutes.
We acknowledge support from CERN and from the national agencies:
CAPES, CNPq, FAPERJ and FINEP (Brazil); 
MOST and NSFC (China); 
CNRS/IN2P3 (France); 
BMBF, DFG and MPG (Germany); 
INFN (Italy); 
NWO (Netherlands); 
MNiSW and NCN (Poland); 
MEN/IFA (Romania); 
MICINN (Spain); 
SNSF and SER (Switzerland); 
NASU (Ukraine); 
STFC (United Kingdom); 
DOE NP and NSF (USA).
We acknowledge the computing resources that are provided by CERN, IN2P3
(France), KIT and DESY (Germany), INFN (Italy), SURF (Netherlands),
PIC (Spain), GridPP (United Kingdom), 
CSCS (Switzerland), IFIN-HH (Romania), CBPF (Brazil),
Polish WLCG  (Poland) and NERSC (USA).
We are indebted to the communities behind the multiple open-source
software packages on which we depend.
Individual groups or members have received support from
ARC and ARDC (Australia);
Minciencias (Colombia);
AvH Foundation (Germany);
EPLANET, Marie Sk\l{}odowska-Curie Actions and ERC (European Union);
A*MIDEX, ANR, IPhU and Labex P2IO, and R\'{e}gion Auvergne-Rh\^{o}ne-Alpes (France);
Key Research Program of Frontier Sciences of CAS, CAS PIFI, CAS CCEPP, 
Fundamental Research Funds for the Central Universities, 
and Sci. \& Tech. Program of Guangzhou (China);
GVA, XuntaGal, GENCAT and Prog.~Atracci\'on Talento, CM (Spain);
SRC (Sweden);
the Leverhulme Trust, the Royal Society
 and UKRI (United Kingdom).

\clearpage
\addcontentsline{toc}{section}{References}
\setboolean{inbibliography}{true}

\ifx\mcitethebibliography\mciteundefinedmacro
\PackageError{LHCb.bst}{mciteplus.sty has not been loaded}
{This bibstyle requires the use of the mciteplus package.}\fi
\providecommand{\href}[2]{#2}

\newpage
\centerline
{\large\bf LHCb collaboration}
\begin
{flushleft}
\small
R.~Aaij$^{32}$\lhcborcid{0000-0003-0533-1952},
A.S.W.~Abdelmotteleb$^{50}$\lhcborcid{0000-0001-7905-0542},
C.~Abellan~Beteta$^{44}$,
F.~Abudin{\'e}n$^{50}$\lhcborcid{0000-0002-6737-3528},
T.~Ackernley$^{54}$\lhcborcid{0000-0002-5951-3498},
B.~Adeva$^{40}$\lhcborcid{0000-0001-9756-3712},
M.~Adinolfi$^{48}$\lhcborcid{0000-0002-1326-1264},
H.~Afsharnia$^{9}$,
C.~Agapopoulou$^{13}$\lhcborcid{0000-0002-2368-0147},
C.A.~Aidala$^{77}$\lhcborcid{0000-0001-9540-4988}
% ...

{\footnotesize \it

% ...
$^{9}$Universit{\'e} Clermont Auvergne, CNRS/IN2P3, LPC, Clermont-Ferrand, France\\
% ...
$^{13}$LPNHE, Sorbonne Universit{\'e}, Paris Diderot Sorbonne Paris Cit{\'e}, CNRS/IN2P3, Paris, France\\
% ...
$^{32}$Nikhef National Institute for Subatomic Physics, Amsterdam, Netherlands\\
% ...
$^{40}$Instituto Galego de F{\'\i}sica de Altas Enerx{\'\i}as (IGFAE), Universidade de Santiago de Compostela, Santiago de Compostela, Spain\\
% ...
$^{44}$Physik-Institut, Universit{\"a}t Z{\"u}rich, Z{\"u}rich, Switzerland\\
% ...
$^{48}$H.H. Wills Physics Laboratory, University of Bristol, Bristol, United Kingdom\\
% ...
$^{50}$Department of Physics, University of Warwick, Coventry, United Kingdom\\
% ...
$^{54}$Oliver Lodge Laboratory, University of Liverpool, Liverpool, United Kingdom\\
% ...
$^{62}$Syracuse University, Syracuse, NY, United States\\
% ...
$^{77}$University of Michigan, Ann Arbor, MI, United States, associated to $^{62}$\\
\bigskip
% ...
}
\end{flushleft}

\end{document}